\documentclass[11pt]{article}

\usepackage{fullpage}
\usepackage{amsfonts,epsfig,graphicx}
\usepackage{amsmath,amssymb,amsthm}
\usepackage{graphics}
\usepackage{macros}
\usepackage[numbers]{natbib}
\usepackage{color}
\usepackage[ruled]{algorithm2e}
\usepackage{epstopdf}
\usepackage{algorithmic}
\usepackage{enumerate}
\usepackage[bookmarks=true,colorlinks,citecolor=blue,urlcolor=blue]{hyperref}

\newtheorem{theorem}{Theorem}
\newtheorem{lemma}{Lemma}

\newtheorem{definition}{Definition}

\makeatletter
\long\def\@makecaption#1#2{
        \vskip 0.8ex
        \setbox\@tempboxa\hbox{\small {\bf #1:} #2}
        \parindent 1.5em  
        \dimen0=\hsize
        \advance\dimen0 by -3em
        \ifdim \wd\@tempboxa >\dimen0
                \hbox to \hsize{
                        \parindent 0em
                        \hfil 
                        \parbox{\dimen0}{\def\baselinestretch{0.96}\small
                                {\bf #1.} #2
                                } 
                        \hfil}
        \else \hbox to \hsize{\hfil \box\@tempboxa \hfil}
        \fi
        }
\makeatother

\newcommand{\ccon}{\ensuremath{c}}
\newcommand{\cconone}{\ensuremath{\ccon_1}}
\newcommand{\ccontwo}{\ensuremath{\ccon_2}}

\newcommand{\nummac}{\ensuremath{m}}

\newcommand{\red}[1]{\textcolor{red}{#1}}

\newcommand{\NUMREP}{\ensuremath{T}}

\newcommand{\Atil}{\ensuremath{\widetilde{A}}}
\newcommand{\Btil}{\ensuremath{\widetilde{B}}}
\newcommand{\matrank}{\ensuremath{r}}
\newcommand{\matdim}{\ensuremath{n}}

\newcommand{\HSPEC}{\ensuremath{H_{\cconone, \ccontwo}}} 

\newcommand{\degpoly}{\ensuremath{p}}
\newcommand{\gaussvec}{\ensuremath{g}}

\newcommand{\FUN}{\ensuremath{F}}
\newcommand{\PROT}{\ensuremath{\Pi}}
\newcommand{\COM}{\mathcal{C}}
\newcommand{\DETCOM}{\mathcal{D}}

\newcommand{\RANDCOME}{\mathcal{R}_\epsilon}

\begin{document} 

\begin{center}

{\LARGE{\bf{{Distributed estimation of generalized matrix rank: \\
Efficient algorithms and lower bounds}}}}

  \vspace{1cm}

  {\large
\begin{tabular}{ccccc}
Yuchen Zhang$^\star$ & & Martin J.\ Wainwright$^{\star, \dagger}$ &&
Michael I. Jordan$^{\star,\dagger}$
\end{tabular}
}

  \vspace{.5cm}

  \texttt{\{yuczhang,wainwrig,jordan\}@berkeley.edu} \\

  \vspace{.5cm}

  {\large $^\star$Department of Electrical Engineering and
  Computer Science
  ~~~~ $^\dagger$Department of Statistics} \\
\vspace{.1cm}

  {\large University of California, Berkeley} \\

  \vspace{.5cm}

\today

  \vspace{.5cm}


\begin{abstract}
We study the following generalized matrix rank estimation problem:
given an $n \times n$ matrix and a constant $c \geq 0$, estimate the
number of eigenvalues that are greater than $c$. In the distributed
setting, the matrix of interest is the sum of $m$ matrices held by
separate machines. We show that any deterministic algorithm solving
this problem must communicate $\Omega(n^2)$ bits, which is
order-equivalent to transmitting the whole matrix. In contrast, we
propose a randomized algorithm that communicates only $\widetilde
\order(n)$ bits. The upper bound is matched by an $\Omega(n)$ lower
bound on the randomized communication complexity. We demonstrate the
practical effectiveness of the proposed algorithm with some numerical
experiments.

\end{abstract}
\end{center}


\section{Introduction}

Given a parameter $\ccon \geq 0$, the generalized rank of an $n \times
n$ positive semidefinite matrix $A$ corresponds to the number of
eigenvalues that are larger than $\ccon$. It is denoted by $\rank(A,
\ccon)$, with the usual rank corresponding to the special case $\ccon
= 0$.  Estimating the generalized rank of a matrix is useful for many
applications. In the context of large-scale principal component analysis
(PCA)~\cite{eckart1936approximation,jolliffe2005principal}, it is
overly expensive to compute the full eigendecomposition before
deciding when to truncate it.  Thus, an important first step is to
estimate the rank of the matrix of interest in order to determine how
many dimensions will be sufficient to describe the data. The rank also
provides useful information for determining the tuning parameter of
robust PCA~\cite{candes2011robust} and collaborative filtering
algorithms~\cite{sarwar2001item,rendle2009bpr}.  In the context of
numerical linear algebra, a number of
eigensolvers~\cite{schofield2012spectrum,polizzi2009density,sakurai2003projection}
for large-scale scientific applications are based on
divide-and-conquer paradigms.  It is a prerequisite of these
algorithms to know the approximate number of eigenvalues located in a
given interval.  Estimating the generalized rank of a matrix is also
needed in the context of sampling-based methods for randomized
numerical linear
algebra~\cite{halko2011finding,mahoney2011randomized}.  For these
methods, the rank of a matrix determines the number of samples
required for a desired approximation accuracy.

Motivated by large-scale data analysis problems, in this paper we 
study the generalized rank estimation problem in a distributed setting, in which 
the matrix $A$ can be decomposed as the the sum of $m$ matrices
\begin{align}
\label{eqn:define-matrix-A}
A \defeq \sum_{i=1}^m A_i,
\end{align}
where each matrix $A_i$ is stored on a separate machine $i$. Thus, a
distributed algorithm needs to communicate between $m$ machines to
perform the estimation.  There are other equivalent formulations of
this problem. For example, suppose that machine $i$ has a design
matrix $X_i \in \R^{n \times N_i}$ and we want to determine the rank of
the aggregated design matrix
\begin{align*}
X & \defeq (X_1, X_2,\ldots, X_m) \in \R^{n\times N} \quad \mbox{where
  $N \defeq \sum_{i=1}^m N_i$}.
\end{align*}
Recall that the singular values of matrix $X$ are equal to
the square root of the eigenvalues of the matrix $XX^T$. If we define
$A_i\defeq X_i X_i^T$, then equation~\eqref{eqn:define-matrix-A}
implies that
\begin{align*}
A = \sum_{i=1}^m A_i = \sum_{i=1}^m X_i X_i^T = X X^T.
\end{align*}
Thus, determining the generalized rank of the matrix $X$ reduces to the
problem of determining the rank of the matrix~$A$.  In this paper, we
focus on the formulation given by
equation~\eqref{eqn:define-matrix-A}.

The standard way of computing the generalized matrix rank, or more
generally of computing the number of eigenvalues within a given
interval, is to exploit Sylvester's law of
inertia~\cite{golub2012matrix}.  Concretely, if the matrix $A - cI$
admits the decomposition $A - cI = L D L^T$, where $L$ is unit lower
triangular and $D$ is diagonal, then the number of eigenvalues of
matrix $A$ that are greater than $c$ is the same as the number of
positive entries in the diagonal of $D$. While this method yields an
exact count, in the distributed setting it requires communicating the
entire matrix $A$.  Due to bandwidth limitations and network
delays, the $\Theta(n^2)$ communication cost is a significant
bottleneck on the algorithmic efficiency. For a matrix of rank $\matrank $,
the power method~\cite{golub2012matrix} can be used to compute the top
$\matrank $ eigenvalues, which reduces the communication cost to $\Theta(\matrank 
n)$.  However, this cost is still prohibitive for moderate sizes of
$\matrank $.  Recently, Napoli et al.~\cite{di2013efficient} studied a more
efficient randomized approach for approximating the eigenvalue counts
based on Chebyshev polynomial approximation of high-pass filters. When
applying this algorithm to the distributed setting, the communication
cost is $\Theta(p n)$, where $p$ is the degree of Chebyshev
polynomials.  However, the authors note that polynomials of high
degree can be necessary.

In this paper, we study the communication complexity of distributed
algorithms for the problem of generalized rank estimation, in both the
deterministic and randomized settings.  We establish upper bounds by
deriving practical, communication-efficient algorithms, and we also
establish complexity-theoretic lower bounds. Our first main result
shows that no deterministic algorithm is efficient in terms of
communication. In particular, communicating $\Omega(n^2)$ bits is
necessary for all deterministic algorithms to approximate the matrix
rank with constant relative error. That such algorithms cannot be 
viewed as efficient is due to the fact that by communicating 
$\order(n^2)$ bits, we are able to compute all eigenvalues and 
the corresponding eigenvectors.  In contrast to the inefficiency 
of deterministic algorithms, we propose a randomized algorithm 
that approximates matrix rank by communicating $\widetilde
\order(n)$ bits. When the matrix is of rank $\matrank $, the relative
approximation error is $1/\sqrt{\matrank }$. Under the same relative error, we
show that $\Omega(n)$ bits of communication is necessary, establishing
the optimality of our algorithm.  This is in contrast with the 
$\Omega(\matrank n)$ communication complexity lower bound for
randomized PCA~\cite{kannan2014principal}. The difference shows that
estimating the eigenvalue count using a randomized algorithm is easier
than estimating the top $\matrank $ eigenpairs.

The research on communication complexity has a long history, dating
back to the seminal work of Yao~\cite{yao1979some} and
Abelson~\cite{abelson1980lower}.  Characterizing the communication
complexity of linear algebraic operations is a fundamental question.
For the problem of rank testing, Chu and
Schnitger~\cite{chu1991communication,chu1995communication} prove the
$\Omega(n^2)$ communication complexity lower bound for
deterministically testing the singularity of integer-valued matrices.
A successful algorithm for this task is required to distinguish two
types of matrices---the singular matrices and the non-singular
matrices with arbitrarily small eigenvalues---a requirement that is
often too severe for practical applications.  Luo and
Tsitsiklis~\cite{luo1993communication} prove an $\Omega(n^2)$ lower
bound for computing one entry of $A^{-1}$, applicable to exact
algorithms (with no form of error allowed).  In contrast, our
deterministic lower bound holds even if we force the non-zero
eigenvalues to be bounded away from zero and allow for approximation
errors, making it more widely applicable to the inexact algorithms
used in practice.  For randomized algorithms, Li et
al.~\cite{sun2012randomized,li2014communication} prove $\Omega(n^2)$
lower bounds for the problems of rank testing, computing a matrix
inverse, and solving a set of linear equations over finite fields.  To
the best of our knowledge, it is not known whether the same lower
bounds hold for matrices in the real field.  In other related work,
Clarkson and Woodruff~\cite{clarkson2009numerical} give an
$\Omega(\matrank ^2)$ space lower bound in the streaming model for
distinguishing between matrices of rank $\matrank$ and $\matrank - 1$.
However, such a space lower bound in the streaming model does not
imply a communication complexity lower bound in the two-way
communication model studied in this paper.


\section{Background and problem formulation}
\label{SecProblem}

In this section, we begin with more details on the problem of
estimating generalized matrix ranks, as well as some background on
communication complexity.

\subsection{Generalized matrix rank}

Given an $n \times n$ positive semidefinite matrix $A$, we use
$\sigma_1(A) \geq \sigma_2(A) \geq \cdots \geq \sigma_n(A) \geq 0$ to
denote its ordered eigenvalues.  For a given constant $\ccon \geq 0$,
the generalized rank of order $\ccon$ is given by
\begin{align}
\rank(A, \ccon) & = \sum_{k=1}^n \Ind[\sigma_k(A) > \ccon],
\end{align}
where $\Ind[\sigma_k(A) > \ccon]$ is a 0-1-valued indicator function
for the event that $\sigma_k(A)$ is larger than $\ccon$.  Since
$\rank(A, 0)$ is equal to the usual rank of a matrix, we see the
motivation for using the generalized rank terminology.  We assume that
$\ltwo{A} = \sigma_1(A) \leq 1$ so that the problem remains on a 
standardized scale.

In an $m$-machine distributed setting, the matrix $A$ can be
decomposed as a sum $A = \sum_{i=1}^m A_i$, where the $n \times n$
matrix $A_i$ is stored on machine $i$.  We study distributed
protocols, to be specified more precisely in the following section, in
which each machine $i$ performs local computation involving the matrix
$A_i$, and the machines then exchange messages so to arrive at an
estimate $\rhat(A) \in [n] \defeq \{0, \ldots, n\}$.  Our goal is to
obtain an estimate that is close to the rank of the matrix in the
sense that
\begin{align}
\label{eqn:esti-desired-bound}
	(1-\delta) \rank(A, \cconone) \leq \rhat(A) \leq (1+\delta)
\rank(A, \ccontwo),
\end{align}
where $\cconone > \ccontwo \geq 0$ and $\delta \in [0, 1)$ are
  user-specified constants. The parameter $\delta \in [0,1)$ upper
    bounds the relative error of the approximation.  The purpose of
    assuming different thresholds $\cconone$ and $\ccontwo$ in
    bound~\eqref{eqn:esti-desired-bound} is to handle the ambiguous
    case when the matrix $A$ has many eigenvalues smaller but very
    close to $\cconone$.  If we were to set $\cconone = \ccontwo$,
    then any estimator $\rhat(A)$ would be strictly prohibited to take
    these eigenvalues into account.  However, since these eigenvalues
    are so close to the threshold, distinguishing them from other
    eigenvalues just above the threshold is obviously difficult (but
    for an uninteresting reason).  Setting $\cconone > \ccontwo$
    allows us to expose the more fundamental sources of difficulty in
    the problem of estimating generalized matrix ranks.  


\subsection{Basics of communication complexity}

To orient the reader, here we provide some very basic background
on communication complexity theory; see the
books~\cite{lee2009lower,KusNis97} for more details.  The standard
set-up in multi-party communication complexity is as follows: suppose
that there are $\nummac$ players (equivalently, agents, machines,
etc.), and for $i \in \{1, \ldots, \nummac\}$, player $i$ holds an
input string $x_i$. In the standard form of communication complexity,
the goal is to compute a joint function $\FUN(x_1, \ldots, x_\nummac)$
of all $\nummac$ input strings with as little communication between
machines as possible.  In this paper, we analyze a communication
scheme known as the \emph{public blackboard model}, in which each
player can write messages on a common blackboard to be read by all
other players.  A distributed protocol $\PROT$ consists of a
coordinated order in which players write messages on the
blackboard. Each message is constructed from the player's local input
and the earlier messages on the blackboard. At the end of the
protocol, some player outputs the value of $\FUN(x_1,\dots,x_\nummac)$
based on the information she collects through the process. The
communication cost of a given protocol~$\Pi$, which we denote by
$\COM(\PROT)$, is the maximum number of bits written on the blackboard
given an arbitrary input.

In a \emph{deterministic protocol}, all messages must be
deterministic functions of the local input and previous messages. The
deterministic communication complexity computing function $\FUN$,
which we denote by $\DETCOM(\FUN)$, is defined by
\begin{align}
\label{EqnDeterministicCC}
\DETCOM(\FUN) & \defeq \min\Big\{ \COM(\PROT):~ \mbox{$\PROT$ is a
  deterministic protocol that correctly computes $\FUN$} \Big\}.
\end{align}
In other words, the quantity $\DETCOM(\FUN)$ is the communication cost
of the most efficient deterministic protocol.

A broader class of protocols are those that allow some form of
randomization. In the public randomness model, each player has access
to an infinite-length random string, and their messages are
constructed from the local input, the earlier messages and the
random string. Let $\mathcal{P}_\epsilon(\FUN)$ be the set of
randomized protocols that correctly compute the function $\FUN$ on any
input with probability at least $1-\epsilon$.  The \emph{randomized
  communication complexity} of computing function $\FUN$ with failure
probability $\epsilon$ is given by
\begin{align}
\label{EqnRandomizedCC}
\RANDCOME(\FUN) & \defeq \min \Big\{ \COM(\PROT) \, \mid \PROT \in
\mathcal{P}_\epsilon(\FUN) \Big\}.
\end{align}

In the current paper, we adopt the bulk of the framework of
communication complexity, but with one minor twist in how we define
``correctness'' in computing the function.  For our problem, each
machine is a player, and the $i^{th}$ player holds the matrix $A_i$.
Our function of interest is given by $\FUN(A_1, \ldots, A_m) = \rank(
\sum_{i=1}^m A_i)$. The public blackboard setting corresponds to a
broadcast-free model, in which each machine can send messages to a
master node, then the master node broadcasts the messages to all other
machines without additional communication cost. 

Let us now clarify the notion of ``correctness'' used in this paper.
In the standard communication model, a protocol $\PROT$ is said to
correctly compute the function $\FUN$ if the output of the protocol is
exactly equal to $\FUN(A_1,\dots,A_\nummac)$. In this paper, we allow
approximation errors in the computation, as specified by the
parameters $(\cconone, \ccontwo)$, which loosen the matrix rank to the
generalized matrix ranks, and the tolerance parameter $\delta \in
(0,1)$.  More specifically, we say:
\begin{definition}
A protocol $\PROT$ correctly computes the rank of the matrix $A$ up to
tolerances $(\cconone, \ccontwo, \delta)$ if the output $\rhat(A)$
satisfies inequality~\eqref{eqn:esti-desired-bound}.  
\end{definition}
Given this definition of correctness, we denote the deterministic
communication complexity of the rank estimation problem by
$\DETCOM(\cconone,\ccontwo,\delta)$, and the corresponding randomized
communication complexity by $\RANDCOME(\cconone,\ccontwo,\delta)$. The
goal of this paper is to study these two quantities, especially their
dependence on the dimension $\matdim$ of matrices.

In addition to allowing for approximation error, our analysis---in
contrast to most classical communication complexity---allows the input
matrices $\{A_i\}_{i=1}^m$ to take real values.  However, doing so
does not make the problem substantially harder.  Indeed, in order to
approximate the matrices in elementwise $\ell_\infty$-norm up to
$\tau$ rounding error, it suffices to discretize each matrix entry
using $\order(\log(1/\tau))$ bits.  As we discuss in more detail in
the sequel, this type of discretization has little effect on the
communication complexity.


\section{Main results and their consequences}
\label{SecMain}

This section is devoted to statements of our main results, as well as
discussion of some of their consequences.


\subsection{Bounds for deterministic algorithms}
\label{SecDeterministic}

We begin by studying the communication complexity of deterministic
algorithms.  Here our main result shows that the trivial
algorithm---the one in which each machine transmits essentially its
whole matrix---is optimal up to logarithmic factors.  In the statement
of the theorem, we assume that the $\matdim$-dimensional matrix $A$ is
known to have rank in the interval\footnote{We use an interval
  assumption, as the problem becomes trivial if the rank is fixed
  exactly.} $[\matrank, 2 \matrank]$ for some integer $\matrank \leq
\matdim/4$.

\begin{theorem}
\label{ThmDet}
For matrices $A$ with rank in the interval $[\matrank, 2 \matrank]$:
\begin{enumerate}
\item[(a)] For all $0 \leq \ccontwo < \cconone$ and $\delta \in
  (0,1)$, we have $\DETCOM(\cconone,\ccontwo,\delta) =
  \order\Big(\nummac \matrank \matdim \log \big(\frac{\nummac \matrank
    \matdim}{\cconone - \ccontwo} \big) \Big)$.
\item[(b)] For two machines $\nummac =2$, constants $0 \leq \ccontwo <
  \cconone < 1/20$ and $\delta \in (0, 1/12)$, we have
  $\DETCOM(\cconone,\ccontwo,\delta) = \Omega(\matrank \matdim)$.
\end{enumerate}
\end{theorem}

When the matrix $A$ has rank $\matrank$ that grows proportionally with
its dimension $\matdim$, the lower bound in part (b) shows that
deterministic communication complexity is surprisingly large: it
scales as $\Theta(n^2)$, which is as large as transmitting the full
matrices.  Up to logarithmic factors, this scaling is matched by the
upper bound in part (a).  It is proved by analyzing an essentially
trivial algorithm: for each index $i = 2, \ldots, \nummac$, machine
$i$ encodes a reduced rank representation of the matrix $A_i$,
representing each matrix entry by $\log_2 \Big(\frac{12 \nummac \matrank
  \matdim}{\cconone - \ccontwo} \Big)$ bits.  It sends this quantized
matrix $\Atil_i$ to the first machine.  Given these
received messages, the first machine then computes the matrix sum
$\Atil \defeq A_1 + \sum_{i=2}^m \Atil_i$, and it outputs $\rhat(A)$
to be the largest integer $k$ such that $\sigma_k(\tildeA) >
(\cconone+\ccontwo)/2$.

On the other hand, in order to prove the lower bound, we consider a
two-party rank testing problem.  Consider two agents holding matrices
$A_1$ and $A_2$, respectively, such that the matrix sum $A \defeq A_1 +
A_2$ has operator norm at most one.  Suppose that exactly one of the
two following conditions are known to hold:
\begin{itemize}
\item the matrix $A$ has rank $\matrank$, or
\item the matrix $A$ has rank between $\frac{6 \matrank}{5}$ and $2
  \matrank$, and in addition its $(6 \matrank/5)^{th}$ eigenvalue is
  lower bounded as $\sigma_{\frac{6 \matrank}{5}}(A) > \frac{1}{20}$.
\end{itemize}
The goal is to decide which case is true by exchanging the minimal
number of bits between the two agents. Denoting this problem by
\ranktest, the proof of part (a) proceeds by showing first that
$D(\ranktest) = \Omega(\matrank \matdim)$, and then reducing from the
\ranktest\ problem to the matrix rank estimation problem.  See
Section~\ref{SecProofThmDet} for the proof.


\subsection{Bounds for randomized algorithms}
\label{SecRandomized}

We now turn to the study of randomized algorithms, for which we see
that the communication complexity is substantially lower.  In
Section~\ref{sec:random-alg}, we propose a randomized algorithm with
$\widetilde \order(n)$ communication cost, and in
Section~\ref{SecRandLower}, we establish a lower bound that matches
this upper bound in various regimes.


\subsubsection{Upper bounds via a practical algorithm}
\label{sec:random-alg}

\begin{figure}[h]
\centering
\includegraphics[width = 0.4\textwidth]{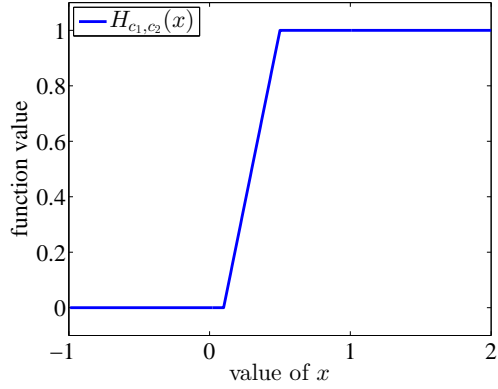}
\caption{An illustration of the function $x \mapsto \HSPEC(x)$ with
  $\cconone = 0.5$ and $\ccontwo = 0.1$.}
\label{fig:hspec}
\end{figure}

In this section, we present an algorithm based on uniform polynomial
approximations for estimating the generalized matrix rank.  Let us
first provide some intuition for the algorithm before defining it more
precisely.  For a fixed pair of scalars $\cconone > \ccontwo \geq 0$,
consider the function $\HSPEC: \R \rightarrow [0,1]$ given by
\begin{align}
\HSPEC(x) & \defeq \begin{cases} 1 & \mbox{if $x > \cconone$} \\ 
0 & \mbox{if $x < \ccontwo$} \\
\frac{x-\ccontwo}{\cconone-\ccontwo} & \mbox{otherwise}.
\end{cases}
\end{align}
As illustrated in Figure~\ref{fig:hspec}, it is a piecewise linear
approximation to a step function.  The squared function $\HSPEC^2$ is
useful in that it can be used to sandwich the generalized ranks of a
matrix $A$.  In particular, given a positive semidefinite matrix $A$
with ordered eigenvalues $\sigma_1(A) \geq \sigma_2(A) \geq \ldots
\geq \sigma_\matdim(A) \geq 0$, observe that we have
\begin{align}
\label{EqnSandwich}
\rank(A,\cconone) \leq \sum_{i=1}^\matdim \HSPEC^2(\sigma_i(A)) \leq
\rank(A,\ccontwo).
\end{align}
Our algorithm exploits this sandwich relation in estimating the
generalized rank.

In particular, suppose that we can find a polynomial function $f: \R
\rightarrow \R$ such that $f \approx \HSPEC$, and which is extended
to a function on the cone of PSD matrices in the standard way.
Observe that if $\sigma$ is an eigenvalue of $A$, then the spectral
mapping theorem~\cite{Bhatia} ensures that $f(\sigma)$ is an
eigenvalue of $f(A)$.  Consequently, letting $\gaussvec \sim
N(0,I_{\matdim \times \matdim})$ be a standard Gaussian vector, we
have the useful relation
\begin{align}
\label{eqn:expec-Ab-expansion}
\E\Big[\ltwos{f(A) \gaussvec}^2\Big] = \sum_{i=1}^n f^2(\sigma_i(A))
\approx \sum_{i=1}^n \HSPEC^2(\sigma_i(A)).
\end{align}
Combined with the sandwich relation~\eqref{EqnSandwich}, we
see that a polynomial approximation $f$ to the function $\HSPEC$ can
be used to estimate the generalized rank.

If $f$ is a polynomial function of degree $\degpoly$, then the vector
$f(A) \gaussvec$ can be computed through $\degpoly$ rounds of
communication.  In more detail, in one round of communciation, we can
first compute the matrix-vector product $A\gaussvec = \sum_{i=1}^m A_i
\gaussvec$.  Given the vector $A\gaussvec$, a second round of
communication suffices to compute the quantity $A^2
\gaussvec$. Iterating a total of $\degpoly$ times, the first machine
is equipped with the collection of vectors $\{\gaussvec, A \gaussvec,
A^2 \gaussvec, \ldots, A^\degpoly \gaussvec \}$, from which it can
compute $f(A) \gaussvec$.

Let us now consider how to obtain a suitable polynomial approximation
of the function $\HSPEC$.  The most natural choice is a Chebyshev
polynomial approximation of the first kind: more precisely, since
$\HSPEC$ is a continuous function with bounded variation, classical
theory~\cite[][Theorem 5.7]{mason2010chebyshev} guarantees that the
Chebyshev expansion converges uniformly to $\HSPEC$ over the interval
$[0,1]$.  Consequently, we may assume that there is a finite-degree
Chebyshev polynomial $q_1$ of the first kind such that
\begin{subequations}
\begin{align}
\label{EqnUniformQ}
\sup_{x \in [0,1]} |q_1(x) - \HSPEC(x)| \leq 0.1.
\end{align}

\begin{figure}[h]
\centering
\begin{tabular}{ccc}
\includegraphics[width = 0.4\textwidth]{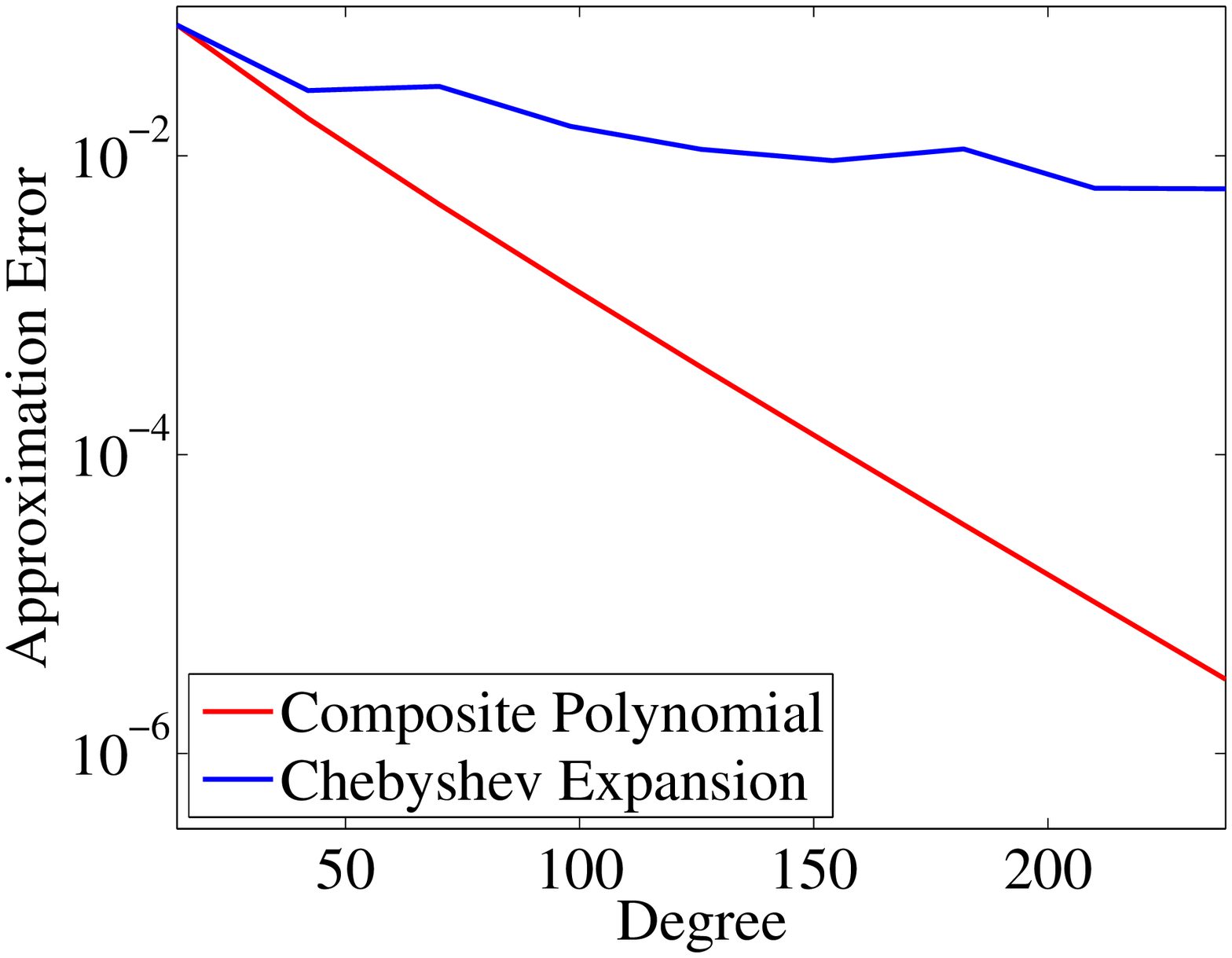} & ~~~ &
\includegraphics[width = 0.4\textwidth]{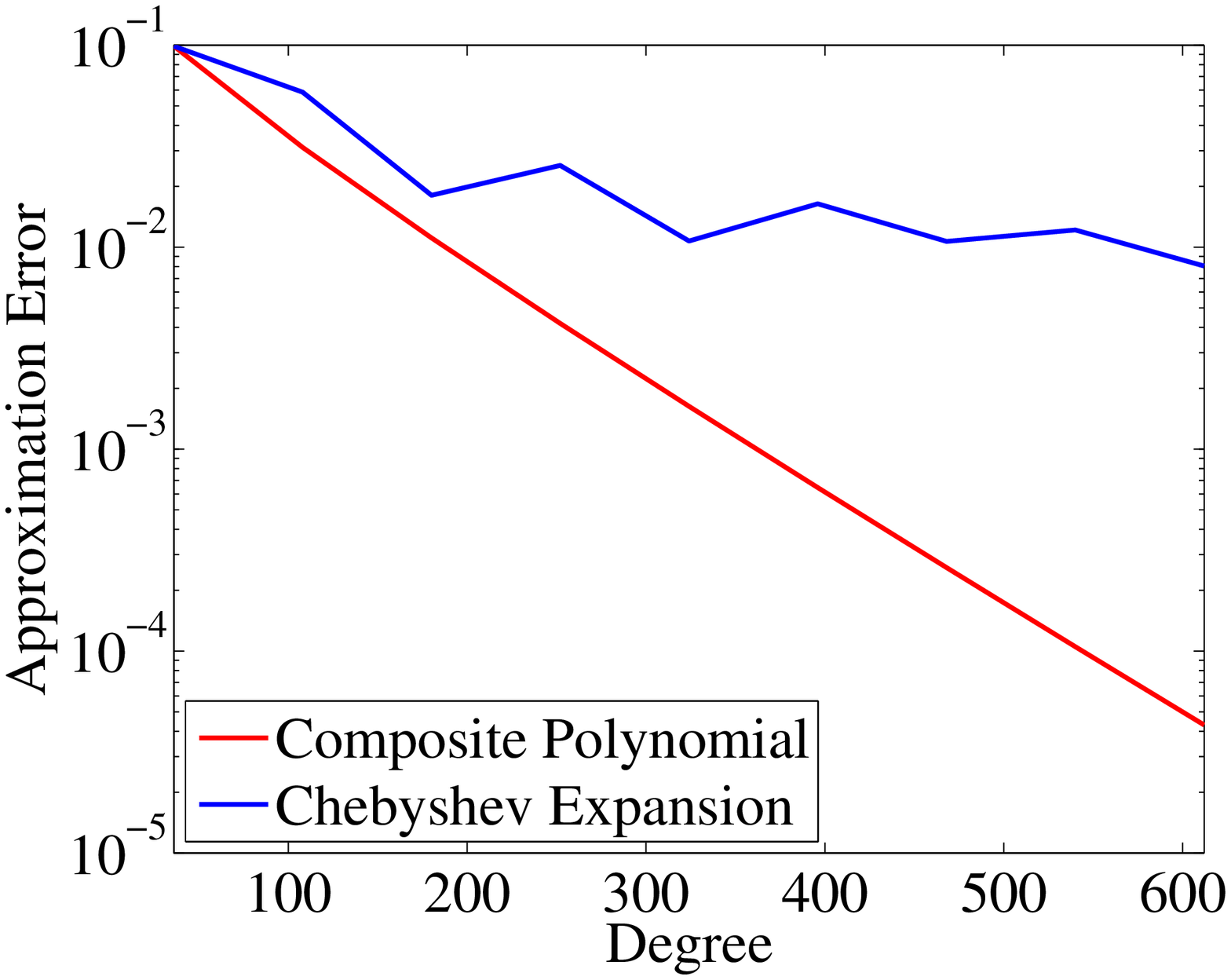}\\ (a)
Thresholds $(\cconone,\ccontwo) = (0.2,0.1)$ & & (b) Thresholds
$(\cconone,\ccontwo) = (0.02,0.01)$
\end{tabular}
\caption{Comparison of the composite polynomial approximation in
  Algorithm~\ref{alg:rand-rank-esti} with the Chebyshev polynomial
  expansion. The error is measured with the $\ell_\infty$-norm on the
  interval $[0,\ccontwo]\cup[\cconone,1]$. The composite polynomial
  approximation achieves a linear convergence rate as the degree is
  increased, while the Chebyshev expansion converges at a much slower
  rate.}
\label{fig:func-error-compare}
\end{figure}

By increasing the degree of the Chebyshev polynomial, we could reduce
the approximation error (set to $0.1$ in the
expansion~\eqref{EqnUniformQ}) to an arbitrarily small level.
However, a very high degree could be necessary to obtain an arbitrary
accuracy. Instead, our strategy is to start with the Chebyshev
polynomial $q_1$ that guarantees the $0.1$-approximation
error~\eqref{EqnUniformQ}, and then construct a second polynomial
$q_2$ such that the composite polynomial function $f = q_2 \circ q_1$
has an approximation error, when measured over the intervals $[0,
  \ccontwo]$ and $[\cconone, 1]$ of interest, that converges linearly
in the degree of function~$f$.  More precisely, consider the
polynomial of degree $2 \degpoly + 1$ given by
\begin{align}
\label{eqn:define-q2}
q_2(x) = \frac{1}{B(\degpoly +1, \degpoly +1)} \int_0^x t^{\degpoly }
(1-t)^{\degpoly} d t \quad \mbox{where $B(\cdot,\cdot)$ is the Beta
  function.}
\end{align}
\end{subequations}

\begin{lemma}
\label{lemma:poly-approx}
Consider the composite polynomial $f(x) \defeq q_2(q_1(x))$, where the
base polynomials $q_1$ and $q_2$ were previously defined in
equations~\eqref{EqnUniformQ} and~\eqref{eqn:define-q2} respectively.
Then $f(x) \in [0,1]$ for all $x \in [0,1]$, and moreover
\begin{align}
\label{EqnTightBound}
|f(x) - \HSPEC(x)| \leq 2^{-\degpoly } \qquad \mbox{for all $x \in
  [0,\ccontwo] \cup [\cconone,1]$.}
\end{align}
\end{lemma}
\noindent See Appendix~\ref{sec:proof-poly-approx} for the proof.

Figure~\ref{fig:func-error-compare} provides a comparison of the error
in approximating $\HSPEC$ for the standard Chebyshev polynomial and
the composite polynomial.  In order to conduct a fair comparison, we
show the approximations obtained by Chebyshev and composite
polynomials of the same final degree, and we evaluate the
$\ell_\infty$-norm approximation error on interval
$[0,\ccontwo]\cup[\cconone,1]$---namely, for a given polynomial
approximation $h$, the quantity
\begin{align*} 
\mbox{Error}(h) \defeq \sup_{x \in [0, \ccontwo] \cup [\cconone, 1]}
|h(x) - \HSPEC(x)|.
\end{align*}
As shown in Figure~\ref{fig:func-error-compare} shows, the composite
polynomial function achieves a linear convergence rate with respect to
its degree. In contrast, the convergence rate of the Chebyshev
expansion is sub-linear, and substantially slower than that of the
composite function. The comparison highlights the advantage of our
approach over the method only based on Chebyshev expansions.

Given the composite polynomial $f = q_2 \circ q_1$, we first evaluate
the vector $f(A) \gaussvec$ in a two-stage procedure. In the first
stage, we evaluate $q_1(A) \gaussvec$, $q_1^2(A) \gaussvec$, $\ldots$,
$q_1^{2p+1}(A) \gaussvec$ using the Clenshaw
recurrence~\cite{clenshaw1955note}, a procedure proven to be
numerically stable~\cite{mason2010chebyshev}.  The details are given
in Algorithm~\ref{alg:poly-eval}. In the second stage, we substitute
the coefficients of $q_2$ so as to evaluate $q_2(q_1(A))b$. The
overall procedure is summarized in Algorithm~\ref{alg:rand-rank-esti}.

\begin{algorithm}[h]
\DontPrintSemicolon \KwIn{$m$ machines hold $A_1,A_2,\dots,A_m\in
  \R^{n\times n}$; vector $v\in \R^d$; Chebyshev polynomial expansion
  $q(x) = \frac{1}{2}a_0 T_0(x) + \sum_{i=1}^d a_i T_i(x)$.}
\KwOut{matrix-vector product $q(A) v$.}
\begin{enumerate}
\item Initialize vector $b_{d+1} = b_{d+2} = {\bf 0}\in \R^n$.
\item For $j=d,\dots,1,0$: the first machine broadcasts $b_{j+1}$ to
  all other machines. Machine $i$ computes $A_i b_{j+1}$ and sends it
  back to the first machine.  The first machine computes
\begin{align*}
b_j \defeq \Big(4\sum_{i=1}^m A_i b_{j+1}\Big) - 2b_{j+1} - b_{j+2} +
a_{j} v.
\end{align*}
\item Output $\frac{1}{2}(a_0 v + b_1 - b_3)$;
\end{enumerate}
\caption{Evaluation of Chebyshev Polynomial}
\label{alg:poly-eval}
\end{algorithm}

\begin{algorithm}[t]
\DontPrintSemicolon \KwIn{Each of $\nummac$ machines hold matrices
  $A_1, A_2, \ldots, A_\nummac \in \R^{\matdim \times
    \matdim}$. Tolerance parameters $(\cconone, \ccontwo)$, polynomial
  degree $\degpoly$, and number of repetitions $\NUMREP$.}
\begin{enumerate}

\item 
\begin{enumerate}[(a)]
\item Find a Chebyshev expansion $q_1$ of the function $\HSPEC$
  satisfying the uniform bound~\eqref{EqnUniformQ}.
\item Define the degree $2 \degpoly + 1$ polynomial function $q_2$ by
  equation~\eqref{eqn:define-q2}.
\end{enumerate}
\item 
\begin{enumerate}[(a)]
\item Generate a random Gaussian vector $\gaussvec \sim N(0,I_{n\times
  n})$.
\item Apply Algorithm~\ref{alg:poly-eval} to compute $q_1(A)
  \gaussvec$, and sequentially apply the same algorithm to compute
  $q_1^2(A) \gaussvec, \dots, q_1^{2p+1}(A) \gaussvec$.
\item Evaluate the vector $y \defeq f(A) \gaussvec = q_2(q_1(A))
  \gaussvec$ on the first machine.
\end{enumerate}
\item Repeat Step 2 for $\NUMREP$ times, obtaining a collection of
  $\matdim$-vectors $\{y_1,\dots,y_ \NUMREP \}$, and output the
  estimate $\rhat(A) = \frac{1}{\NUMREP}\sum_{i=1}^\NUMREP
  \ltwos{y_i}^2$.

\end{enumerate}
\label{alg:rand-rank-esti}
\caption{Randomized Algorithm for Rank Estimation}
\end{algorithm}

The following result provides a guarantee for the overall
procedure (combination of Algorithm~\ref{alg:poly-eval} and 
Algorithm~\ref{alg:rand-rank-esti}) when run with
degree $\degpoly = \lceil \log_2(2 \matdim) \rceil$:
\begin{theorem}
\label{ThmRandUpper}
For any $0 \leq \delta < 1$, with probability at least $1 - 2 \exp\left( -\frac{
    \NUMREP \delta^2 \rank(A,\cconone)}{32}\right)$, the output of
  Algorithm~\ref{alg:rand-rank-esti} satisfies the bounds
\begin{align}
\label{eqn:rand-alg-bound}
(1-\delta)\rank(A,\cconone) - 1 \leq \rhat(A) \leq
(1+\delta)(\rank(A,\ccontwo)+1).
\end{align}
Moreover, we have the following upper bound on
the randomized communication complexity of estimating the
generalized matrix rank:
\begin{align}
\label{EqnRandUpper}
\RANDCOME \Big(\cconone,\ccontwo,1/\sqrt{\rank(A,\cconone)}\Big) =
\widetilde \order(\nummac \matdim).
\end{align}
\end{theorem}
\noindent 
We show in Section~\ref{SecRandLower} that the upper
bound~\eqref{EqnRandUpper} is unimprovable up to the logarithmic
pre-factors.  For now, let us turn to the results of some numerical
experiments using Algorithm~\ref{alg:rand-rank-esti}, which show that in
addition to being an order-optimal algorithm, it is also practically
useful.


\subsubsection{Numerical experiments}
\label{SecNumerical}

Given $\nummac = 2$ machines, suppose that machine $i$ (for $i = 1,
2$) receives $N_i = 1000$ data points of dimension $n = 1000$.  Each
data point $x$ is independently generated as $x = a + \varepsilon$,
where $a\sim N(0, \lambda\Sigma)$ and $\varepsilon \sim N(0, \sigma^2
I_{n\times n})$ are random Gaussian vectors.  Here $\Sigma\in
\R^{n\times n}$ is a low-rank covariance matrix of the form $\Sigma
\defeq \sum_{i=1}^\matrank u_i u_i^T$, where $\{u_i\}_{i=1}^\matrank$
are an orthonormal set of vectors in $\real^\matdim$ drawn uniformly
at random.  The goal is to estimate the rank $\matrank$ from the
observed $N_1 + N_2 = 2000$ data points.

\begin{figure}
\centering
\begin{tabular}{ccc}
\includegraphics[width = 0.4\textwidth]{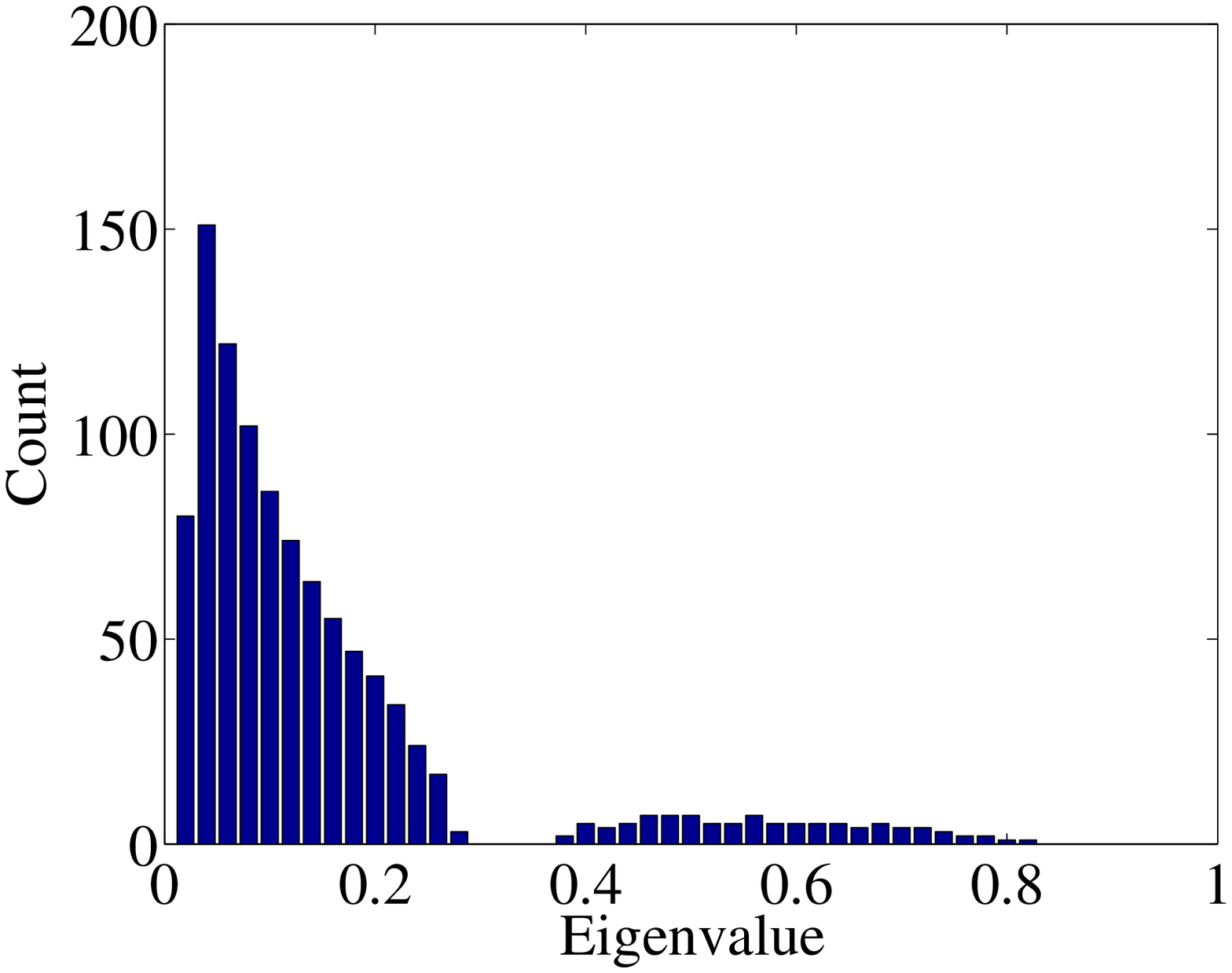} & ~~~
& \includegraphics[width = 0.4\textwidth]{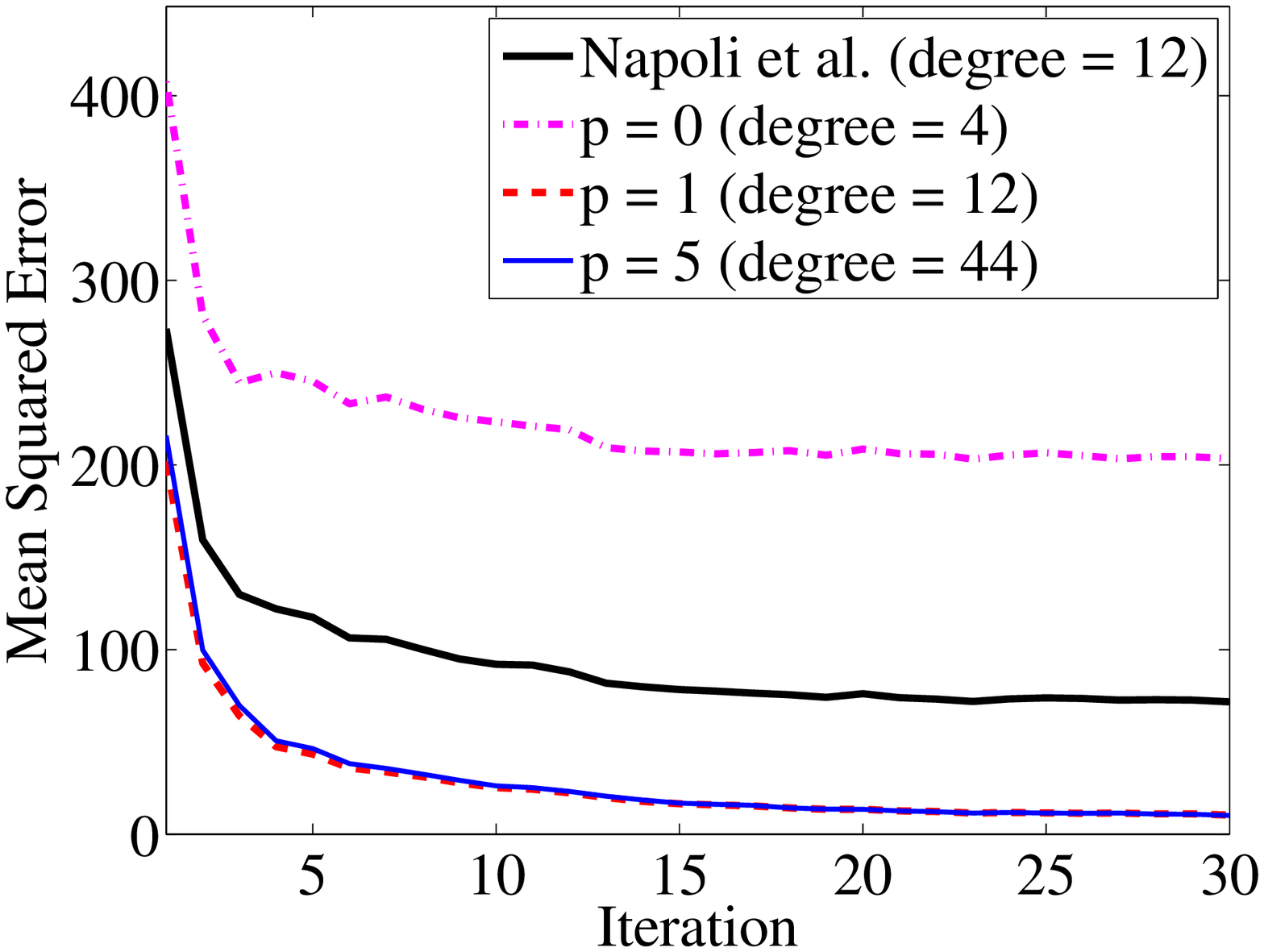}\\
(a) Eigenvalue Distribution & & (b) Rank Estimation Error
\end{tabular}
\caption{Panel (a): distribution of eigenvalues of matrix $A$.  Panel
  (b): mean squared error of rank estimation versus the number of
  iterations for the baseline method by Napoli et
  al.~\cite{di2013efficient}, and three versions of
  Algorithm~\ref{alg:rand-rank-esti} (with parameters $p \in
  \{0,1,5\}$).}
\label{fig:rank-error}
\end{figure}

Let us now describe how to estimate the rank using the covariance
matrix of the samples. Notice that $\E[xx^T] = \lambda^2\Sigma +
\sigma^2 I_{n\times n}$, of which there are $r$ eigenvalues equal to
$\lambda + \sigma^2$ and the remaining eigenvalues are equal to
$\sigma^2$.  Letting $x_{i,j} \in \R^\matdim$ denote the $j$-th data
point received by machine $i$, that machine can compute the local
sample covariance matrix
\begin{align*}
A_i = \frac{1}{N_1+N_2}\sum_{j=1}^{N_i} x_{i,j} x_{i,j}^T, \quad
\mbox{for $i = 1, 2$.}
\end{align*}
The full sample covariance matrix is given by the sum $A \defeq A_1 +
A_2$, and its rank can be estimated using
Algorithm~\ref{alg:rand-rank-esti}.

In order to generate the data, we choose the parameters $\matrank =
100$, $\lambda = 0.4$ and $\sigma^2 = 0.1$.  These choices motivate
the thresholds $\cconone = \lambda + \sigma^2 = 0.5$ and $\ccontwo =
\sigma^2 = 0.1$ in Algorithm~\ref{alg:rand-rank-esti}.  We illustrate
the behavior of the algorithm for three different choices of the
degree parameter $\degpoly$---specifically, $\degpoly \in
\{0,1,5\}$---and for a range of repetitions $\NUMREP \in \{1,2,
\ldots, 30 \}$.  Letting $\rhat(A)$ denote the output of the
algorithm, we evaluate the mean squared error, $\E[(\rhat(A) -
r)^2]$, based on $100$ independent runs of the algorithm.

We plot the results of this experiment in Figure~\ref{fig:rank-error}.
Panel (a) shows the distribution of eigenvalues of the matrix $A$. In
this plot, there is a gap between the large eigenvalues generated by
the low-rank covariance matrix $\Sigma$, and small eigenvalues
generated by the random Gaussian noise, showing that the problem is
relatively easy to solve in the centralized setting. Panel (b) shows
the estimation error achieved by the communication-efficient
distributed algorithm; notice how the estimation error stabilizes
after $\NUMREP = 30$ repetitions or iterations. We compare our
algorithm for $\degpoly \in \{0,1,5\}$, corresponding to polynomial
approximations with degree in $\{4,12,44\}$. For the case $\degpoly =
0$, the polynomial approximation is implemented by the Chebyshev
expansion. For the case $\degpoly = 1$ and $\degpoly = 5$, the
approximation is achieved by the composite function $f$.  As a
baseline method, we also implement Napoli et al.'s
algorithm~\cite{di2013efficient} in the distributed setting. In
particular, their method replaces the function $f$ in
Algorithm~\ref{alg:rand-rank-esti} by a Chebyshev expansion of the
high-pass filter $\indicator(x \geq \frac{\cconone+\ccontwo}{2})$.  It
is observed that both the Chebyshev expansion with $\degpoly = 0$ and
the baseline method incur a large bias in the rank estimate, while
the composite function's estimation errors are substantially
smaller. After $\NUMREP = 30$ iterations,
Algorithm~\ref{alg:rand-rank-esti} with $\degpoly = 1$ achieves a mean
squared error close to $10$, which means that the relative error of
the estimation is around~$3\%$.


\subsubsection{Lower Bound}
\label{SecRandLower}

It is natural to wonder if the communication efficiency of
Algorithm~\ref{alg:rand-rank-esti} is optimal. The following theorem
shows that, in order to achieve the same $1/\sqrt{\matrank}$ relative
error, it is necessary to send $\Omega(\matdim)$ bits.  As in our
upper bound, we assume that the matrix $A$ satisfies the spectral norm
bound $\ltwos{A} \leq 1$.  Given an arbirary integer $\matrank$ in the
interval $[16, \matdim/4]$, suppose that the generalized matrix ranks
satisfy the sandwich relation $\matrank \leq \rank(A,\cconone) \leq
\rank(A,\ccontwo) \leq 2 \matrank$.  Under these conditions, we have
the following guarantee:
\begin{theorem}
\label{ThmRandLower}
For any $\cconone,\ccontwo$ satisfying $\cconone < 2\ccontwo \leq 1$
and any $\epsilon \leq \epsilon_0$ for some numerical constant~$\epsilon_0$, we have
\begin{align}
\label{eqn:random-lower-expr}
\RANDCOME \Big(\cconone,\ccontwo,1/\sqrt{\matrank}\Big) =
\Omega(\matdim).
\end{align}
\end{theorem}
\noindent See Section~\ref{SecProofThmRandLower} for the proof of this
lower bound. \\

According to Theorem~\ref{ThmRandLower}, for matrices with true rank
in the interval $[16, \matdim/2]$, the communication
complexity for estimating the rank with relative error
$1/\sqrt{\matrank}$ is lower bounded by $\Omega(\matdim)$.  This lower
bound matches the upper bound provided by Theorem~\ref{ThmRandUpper}.
In particular, choosing $\matrank = 16$ yields the worst-case lower
bound
\begin{align*}
\RANDCOME (\cconone,\ccontwo,1/4) = \Omega(\matdim),
\end{align*}
showing that $\Omega(\matdim)$ bits of communication are necessary for
achieving a constant relative error.  This lower bound is not trivial
relative to the coding length of the correct answer: given that the
matrix rank is known to be between $\matrank$ and $2 \matrank$, this
coding length scales only as $\Omega(\log \matrank)$.

There are several open problems suggested by the result of
Theorem~\ref{ThmRandLower}.  First, it would be interesting to
strengthen the lower bound~\eqref{eqn:random-lower-expr} from
$\Omega(\matdim)$ to $\Omega(\nummac \matdim)$, incorporating the
natural scaling with the number of machines $\nummac$. Doing so
requires a deeper investigation into the multi-party structure of the problem.
Another open problem is to lower bound the communication complexity
for arbitrary values of the tolerance parameter $\delta$, say as small
as $1/\matrank $.  When $\delta$ is very small, communicating
$\order(\nummac \numobs^2)$ bits is an obvious upper bound, and we are
not currently aware of better upper bounds. On the other hand, whether it is possible to prove an
$\Omega(\matdim^2)$ lower bound for small~$\delta$ remains an open
question.


\section{Proofs}

In this section, we provide the proofs of our main results, with the
proofs of some more technical lemmas deferred to the appendices.


\subsection{Proof of Theorem~\ref{ThmDet}}
\label{SecProofThmDet}

Let us begin with our first main result on the deterministic
communication complexity of the generalized rank problem.


\subsubsection{Proof of lower bound}

We first prove the lower bound stated in part (a) of
Theorem~\ref{ThmDet}.  Let us recall the \ranktest\ problem previously
described after the statement of Theorem~\ref{ThmDet}.  Alice holds a
matrix $A_1\in \R^{n\times n}$ and Bob holds a matrix $A_2\in
\R^{n\times n}$ such that the matrix sum $A \defeq A_1 + A_2$ has
operator norm at most one.  Either the matrix $A$ has rank $\matrank$,
or the matrix $A$ has rank between $\frac{6 \matrank}{5}$ and
$2\matrank $, and in addition its $6 \matrank/5$ eigenvalue is lower
bounded as $\sigma_{\frac{6 \matrank}{5}}(A) > \frac{1}{20}$.  The
\ranktest\ problem is to decide which of these two mutually exclusive
alternatives holds.  The following lemma provides a lower bound on the
deterministic communication complexity of this problem:

\begin{lemma}
\label{LemRankTest}
For any $\matrank \leq n/4$, we have $D(\ranktest) = \Omega(\matrank
n)$.
\end{lemma}

We use Lemma~\ref{LemRankTest} to lower bound
$D(\cconone,\ccontwo,\delta)$, in particular by reducing to it from
the \ranktest\ problem. Given a \ranktest\ instance, since there are
$m \geq 2$ machines, the first two machines can simulate Alice and
Bob, holding $A_1$ and $A_2$ respectively. All other machines hold a
zero matrix.  Suppose that $\cconone \leq 1/20$ and $\delta \leq
1/12$.  If there is an algorithm achieving the
bound~\eqref{eqn:esti-desired-bound}, then if $A = A_1 + A_2$ is of
rank $\matrank$, then
\begin{subequations}
\begin{align}
\label{eqn:rhat-A-deter-upper}
\rhat(A) \leq (1+\delta)\rank(A,\ccontwo) \leq \Big( 1 + \frac{1}{12}
\Big) \matrank = \frac{13 \matrank}{12}.
\end{align}
Otherwise, the $\frac{6 \matrank}{5}$-th eigenvalue of $A$ is greater
than $1/20$, so that
\begin{align}
\label{eqn:rhat-A-deter-lower}
\rhat(A) \geq (1-\delta)\rank(A,\cconone) \geq \Big( 1 - \frac{1}{12}
\Big) \frac{6 \matrank}{5} = \frac{11 \matrank}{10} > \frac{13
  \matrank}{12}.
\end{align}
\end{subequations}
In conjunction, inequality~\eqref{eqn:rhat-A-deter-upper}
and~\eqref{eqn:rhat-A-deter-lower} show that we can solve the
\ranktest\ problem by testing whether or not $\rhat(A) \leq \frac{13
  \matrank}{12}$.  Consequently, the deterministic communication
complexity $D(\cconone,\ccontwo,\delta)$ is lower bounded by the
communication complexity of \ranktest.\\

In order to complete the proof of Theorem~\ref{ThmDet}(a), it remains
to prove Lemma~\ref{LemRankTest}, and we do so using a randomized
construction.  Let us say that a matrix $Q \in \real^{\matrank \times
\matdim}$ is sampled from the orthogonal ensemble if it is sampled
in the following way: let $U\in \real^{\matdim\times \matdim}$ be a matrix
uniformly sampled from the group of orthogonal matrices, then
$Q$ is the sub-matrix consisting of the first $\matrank$ rows of $U$.
We have the following claim.

\begin{lemma}
\label{lemma:intersec-ortho-space}
Given matrices $Q_1\in
\R^{\matrank \times \matdim}$ and $Q_2 \in \R^{\matrank \times
  \matdim}$ independently sampled from the orthogonal ensemble, we have $\sigma_{\frac{6\matrank}{5}}(Q_1^T Q_1 + Q_2^T
Q_2) > \frac{1}{10}$ with probability at least $1 - e^{- \frac{3
    \matrank \matdim}{100}}$.
\end{lemma}
\noindent See Appendix~\ref{sec:proof-intersec-ortho-space} for the
proof. \\

Taking Lemma~\ref{lemma:intersec-ortho-space} as given, introduce the
shorthand $N = \lfloor \frac{\matrank \matdim}{50} \rfloor$.  Suppose
that we independently sample $2^N$ matrices of dimensions
$\matrank \times \matdim$ from the orthogonal ensemble.  Since there are $2^{N}(2^N-1)/2$ distinct
pairs of matrices in our sample, the union bound in conjunction with
Lemma~\ref{lemma:intersec-ortho-space} implies that
\begin{align}
\label{eqn:union-ortho-rank-bound}
\prob\Big[ \forall i\neq j:~ \sigma_{\frac{6\matrank}{5}}(Q_i^T Q_i +
  Q_j^T Q_j) > \frac{1}{10} \Big] \geq 1 - \frac{2^{N}(2^N-1)}{2}
\exp\Big(- \frac{3 \matrank \matdim}{100}\Big).
\end{align}
With our choice of $N$, it can be verified that the right-hand side of
inequality~\eqref{eqn:union-ortho-rank-bound} is positive. Thus, there
exists a realization of orthogonal matrices $Q_1,\dots,Q_{2^N}\in
\R^{\matrank \times \matdim}$ such that for all $i\neq j$ we have
$\sigma_{\frac{6 \matrank}{5}}(Q_i^T Q_i + Q_j^T Q_j) > \frac{1}{10}$.

We use this collection of orthogonal matrices in order to reduce the
classical \equality\ problem to the rank estimation problem.  In the
\equality\ problem, Alice has a binary string $x_1\in\{0,1\}^N$ and
Bob has another binary string $x_2\in \{0,1\}^N$, and their goal is to
compute the function
\begin{align*}
\equality(x_1,x_2) & =
\begin{cases} 1  & \mbox{if $x_1=x_2$;} \\ 
0 & \mbox{otherwise;}
\end{cases}
\end{align*}
It is well-known~\cite{KusNis97} that the deterministic communication
complexity of the \equality\ problem is $D(\equality) = N+1$.

In order to perform the reduction, given binary strings $x_1$ and
$x_2$ of length $N$, we construct two matrices $A_1$ and $A_2$ such
that their sum $A = A_1 + A_2$ has rank $\matrank$ if and only if
$x_1=x_2$.  Since both $x_1$ and $x_2$ are of length $N$, each of them
encodes an integer between $1$ and $2^N$.  Defining $A_1 =
\frac{Q_{x_1}^T Q_{x_1}}{2}$ and $A_2 = \frac{Q_{x_2}^T Q_{x_2}}{2}$,
the triangle inequality guarantees that
\begin{align*}
\ltwos{A} \leq \ltwos{A_1} + \ltwos{A_2} = \frac{\ltwos{Q_{x_1}^T
    Q_{x_1}} + \ltwos{Q_{x_2}^T Q_{x_2}}}{2} \leq 1,
\end{align*}
showing that $A$ satisfies the required operator norm bound.  If $x_1
= x_2$, then $A = Q_{x_1}^T Q_{x_1}$, which is a matrix of rank
$\matrank$. If $x_1 \neq x_2$, then by our construction of $Q_{x_1}$
and $Q_{x_2}$, we know that the matrix $A$ has rank between
$\frac{6\matrank}{5}$ and $2 \matrank$ and moreover that
$\sigma_{\frac{6\matrank}{5}}(A) > \frac{1}{20}$.  Thus, we can output
$\equality(x_1,x_2)=1$ if we detect the rank of matrix $A$ to be
$\matrank$ and output $\equality(x_1,x_2)=0$ otherwise.  Using this
protocol, the \equality\ evaluation is always correct. As a
consequence, the deterministic communication complexity of
\ranktest\ is lower bounded by that of \equality.  Finally, noting
that \mbox{$D(\ranktest) \geq D(\equality) = N + 1 > \frac{\matrank
    \matdim}{50}$} completes the proof.


\subsubsection{Proof of upper bound}

In order to prove the upper bound stated in part (b), we analyze the
algorithm described following the theorem statement.  If the matrix $A
= \sum_{i=1}^\nummac A_i$ has rank at most $2 \matrank$, then given the
PSD nature of the component matrices, each matrix $A_i$ also has rank
at most $2 \matrank$.  Consequently, we can find a factorization of the
form $A_i = B_i B_i^T$ where $B_i \in \real^{\matdim \times
  \matrank}$.  Let $\Btil_i$ be a quantization of the matrix $B_i$,
allocating $\log_2 \big( \frac{12 \nummac \matrank \matdim}{\cconone -
  \ccontwo} \big)$ bits to each entry.  Note that each machine must
transmit at most $\matrank \matdim \log_2 \Big( \frac{12 \nummac
  \matrank \matdim}{\cconone - \ccontwo} \Big)$ bits in order to
convey the quantized matrix $\Btil_i$.

Let us now analyze the approximation error.  By our choice of
quantization, we have
\begin{align*}
\opnorm{\Btil_i - B_i} & \leq \frobnorm{\Btil_i - B_i} \; \leq \;
\sqrt{2 \matrank \matdim} \frac{\cconone - \ccontwo}{12 \nummac \matrank
  \matdim} \; = \; \frac{\cconone - \ccontwo}{6 \nummac \,
  \sqrt{2 \matrank \matdim}}.
\end{align*}
Defining $\Atil_i = \Btil_i \Btil_i^T$ we have
\begin{align*}
\frobnorm{\Atil_i - A_i} \leq \frobnorm{\Btil_i - B_i} \,
\sqrt{2 \matrank \matdim} \Big( \opnorm{B_i} + \opnorm{\Btil_i} \Big) &
\leq \frac{\cconone - \ccontwo}{6 \nummac} \Big( 2 + \frac{\cconone -
  \ccontwo}{6 \nummac \, \sqrt{2 \matrank \matdim}} \Big) \\
& \leq \frac{\cconone - \ccontwo}{2 \nummac},
\end{align*}
where the final inequality follows as long as $ \frac{\cconone -
  \ccontwo}{6 \nummac \, \sqrt{2 \matrank \matdim}} \leq 1$.

Consequently, the sum $\Atil = \sum_{i=1}^m \Atil_i$ satisfies the
bound 
\begin{align*}
\norms{\Atil- A}_F \leq \sum_{i=2}^m \norms{\Atil_i - A_i}_F \leq
\frac{(\cconone-\ccontwo)}{2}.
\end{align*}
Applying the Wielandt-Hoffman inequality~\cite{Horn85} yields the
upper bound
\begin{align}
\label{eqn:matrix-eigen-perturb-bound}
| \sigma_k(\tildeA) - \sigma_k(A) | \leq \norms{\tildeA - A}_F \leq
(\cconone-\ccontwo)/2 \quad \mbox{for all $k\in[n]$.}
\end{align}
Recalling that $\rhat(A)$ is the largest integer $k$ such that
$\sigma_k(\tildeA) > (\cconone+\ccontwo)/2$, inequality
~\eqref{eqn:matrix-eigen-perturb-bound} implies that
\begin{align*}
(\cconone+\ccontwo)/2 \geq \sigma_{\rhat(A)+1}(\tildeA) \geq
  \sigma_{\rhat(A)+1}(A) - (\cconone-\ccontwo)/2,
\end{align*}
which implies $\sigma_{\rhat(A)+1}(A) \leq \cconone$. This upper bound
verifies that $\rhat(A) \geq \rhat(A,\cconone)$. On the other hand,
inequality~\eqref{eqn:matrix-eigen-perturb-bound} also yields
\begin{align*}
(\cconone+\ccontwo)/2 < \sigma_{\rhat(A)}(\tildeA) \leq \sigma_{\rhat(A)}(A) +
  (\cconone-\ccontwo)/2,
\end{align*}
which implies $\sigma_{\rhat(A)}(A) > \ccontwo$ and $\rhat(A) \leq
\rhat(A,\ccontwo)$. Combining the above two inequalities yields the
claim~\eqref{eqn:esti-desired-bound}.


\subsection{Proof of Theorem~\ref{ThmRandUpper}}
\label{SecProofThmRandUpper}

We split the proof into two parts, corresponding to the upper
bounds~\eqref{eqn:rand-alg-bound} and~\eqref{EqnRandUpper}
respectively.

\paragraph{Proof of upper bound~\eqref{eqn:rand-alg-bound}:}

Let $\lambda_j$ be the $j$-th largest eigenvalue of $A$ and let $v_j$
be the associated eigenvector.  Let function $f$ be defined as $f(x)
\defeq q_2(q_1(x))$. Using basic linear algebra, we have
\begin{align}
\label{eqn:closed-form-y}
\ltwos{y}^2 = \sum_{j=1}^n f^2(\lambda_j) (v_j^T \gaussvec)^2.
\end{align}
Since $\gaussvec$ is an isotropic Gaussian random vector, the random
variables $Z_j = (v_j^T \gaussvec)^2$ are i.i.d., each with
$\chi^2$ distribution with one degree of freedom. To analyze 
the concentration behavior of $Z$ variables, we recall the
notion of a sub-exponential random variable.

A random variable $Y$ is called \emph{sub-exponential} with parameter
$(\sigma^2,\beta)$ if $\E[Y] = 0$ and the moment generating function
is upper bounded as \mbox{$\E[e^{tY}] \leq e^{t^2\sigma^2/2}$} for all
$|t|\leq 1/\beta$.  The following lemma, proved in
Appendix~\ref{sec:proof-sub-exponential}, characterizes some basic
properties of sub-exponential random variables.
\begin{lemma}
  \label{lemma:sub-exponential}
\begin{enumerate}[(a)]
\item If $Z \sim \chi^2$, then both $Z-1$ and $1-Z$ are
  sub-exponential with parameter $(4,4)$.
\item Given an independent sequence $\{Y_i\}_{i=1}^\matdim$ in which
  $Y_i$ is sub-exponential with parameter $(\sigma^2_i,\beta_i)$, then
  for any choice of non-negative weights $\{\alpha\}_{i=1}^\matdim$,
  the weighted sum $\sum_{i=1}^n \alpha_i Y_i$ is sub-exponential with
  parameters $(\sum_{i=1}^n
  \alpha_i^2\sigma_i^2,\max_{i\in[n]}\{\alpha_i\beta_i\})$.
\item If $Y$ is sub-exponential with parameter $(\sigma^2,\beta)$,
  then
\begin{align*}
\mprob \big[ Y \geq t \big] \leq e^{ -\frac{t^2}{2\sigma^2}} \quad
\mbox{for all $t \in [0, \frac{\sigma^2}{\beta})$.}
\end{align*}
\end{enumerate}
\end{lemma}

\noindent
We consider $\ltwos{y}^2$ as well as the associated lower bound $L =
\sum_{j=1}^{ \rank(A,\cconone)} f^2(\lambda_j) (v_j^T b)^2$.  By parts
(a) and (b) of Lemma~\ref{lemma:sub-exponential}, the variable
$\ltwos{y}^2 - \E[\ltwos{y}^2]$ is sub-exponential with parameter
$(4\sum_{i=1}^n f^2(\lambda_i),4)$, and the variable $\E[L] - L$ is
sub-exponential with parameter $(4\sum_{i=1}^{\rank(A,\cconone)}
f^2(\lambda_i),4)$. In order to apply part (c) of
Lemma~\ref{lemma:sub-exponential}, we need upper bounds on the sum
$\sum_{i=1}^n f^2(\lambda_i)$, as well as upper/lower bounds on the
sum $\sum_{i=1}^{\rank(A,\cconone)} f^2(\lambda_i)$. For the first
sum, we have
\begin{align}
\sum_{j=1}^n f^2(\lambda_j) &= \sum_{j=1}^{ \rank(A,\ccontwo)}
f^2(\lambda_j) + \sum_{j= \rank(A,\ccontwo)+1}^{n}
f^2(\lambda_j)\nonumber\\ & \leq \rank(A,\ccontwo) + \matdim
2^{-\degpoly} \nonumber \\
\label{eqn:sum-f-bound}
& \leq \rank(A,\ccontwo) + 1.
\end{align}
where the last two inequalities use Lemma~\ref{lemma:poly-approx} and
the fact that $p = \lceil \log_2(2n) \rceil$.  For the second sum,
using Lemma~\ref{lemma:poly-approx} implies that 
\begin{align*}\rank(A,\cconone)
&\geq \sum_{i=1}^{\rank(A,\cconone)} f^2(\lambda_i) \geq
\rank(A,\cconone) (1 - 2^{-p})^2 \\
&\stackrel{(i)}{\geq} \rank(A,\cconone) (1 - 1/(2n))^2 \stackrel{(ii)}{\geq} \rank(A,\cconone) - 1.  
\end{align*}
where inequality (i) follows since $2^{-p} \leq 1/(2n)$; inequality (ii) follows since
$(1-1/(2n))^2 \geq 1 - 1/n$.
Thus, we have
\begin{align}
\E[\ltwos{y}^2] \leq \rank(A,\ccontwo) + 1 \quad \mbox{and} \quad
\E[L] \geq \rank(A,\cconone) - 1.\label{eqn:expectation-bounds}
\end{align}
Putting together the pieces, we see that $\ltwos{y}^2 -
\E[\ltwos{y}^2]$ is sub-exponential with parameter
$(4(\rank(A,\ccontwo) + 1),4)$ and $\E[L] - L$ is sub-exponential with
parameter $(4\;\rank(A,\cconone),4)$.

Let $\rhat$ be the average of $\NUMREP$ independent copies of
$\ltwos{y}$, and let $\rhat_L$ be the average of $\NUMREP$ independent
copies of $L$. By Lemma~\ref{lemma:sub-exponential} (b), we know that
$\rhat-\E[\rhat]$ is sub-exponential with parameter $(4(
\rank(A,\ccontwo) + 1)/\NUMREP,4/\NUMREP)$, and $\E[\rhat_L] -
\rhat_L$ is sub-exponential with parameter
$(4\;\rank(A,\cconone)/\NUMREP,4/\NUMREP)$.  Plugging these parameters
into Lemma~\ref{lemma:sub-exponential} (c), for any $0\leq \delta <
1$, we find that
\begin{subequations}
\begin{align}
\prob\Big[ \rhat \leq \E[\rhat] + \delta(\rank(A,\ccontwo)+1) \Big] &
\geq 1- \exp\Big( -\frac{ \NUMREP \delta^2 ( \rank(A,\ccontwo)+1)}{32}\Big)
	\label{eqn:S-upper-bound} \\
\prob \Big[ \rhat_L \geq \E[\rhat_L] - \delta \rank(A,\cconone) \Big]
& \geq 1- \exp \Big( -\frac{ \NUMREP \delta^2 \rank(A,\cconone)}{32}\Big).
\label{eqn:S-lower-bound}
\end{align}
\end{subequations}
Combining inequalities~\eqref{eqn:expectation-bounds},
~\eqref{eqn:S-upper-bound}, and~\eqref{eqn:S-lower-bound} yields
\begin{align}
\label{eqn:S-lower-upper}
\prob\Big[ (1-\delta) \rank(A,\cconone) - 1 \leq \rhat_L \leq \rhat
  \leq (1+\delta)( \rank(A,\ccontwo)+1) \Big] \leq 1 - 2 e^{ -\frac{
    \NUMREP \delta^2 \rank(A,\cconone)}{32}},
\end{align}
which completes the proof of inequality~\eqref{eqn:rand-alg-bound}. \\

\paragraph{Proof of upper bound~\eqref{EqnRandUpper}:}
It remains to show to establish the upper bound~\eqref{EqnRandUpper}
on the randomized communication complexity.  The subtle issue is that
in a discrete message model, we cannot calculate $f(A) \gaussvec$ without
rounding errors. Indeed, in order to make the rounding error of each
individual message bounded by $\tau$, each machine needs
$\order(\numobs \log(1/\tau))$ bits to encode a message. Consequently,
the overall communication complexity scales as $\order( \NUMREP
\nummac d \degpoly \matdim \log(1/\tau))$, where $\NUMREP$ is the
number of iterations of Algorithm~\ref{alg:rand-rank-esti}; $m$ is the
number of machines, the quantities $d$ and $p$ are the degrees of
$q_1$ and $q_2$, and $n$ is the matrix dimenson.  With the choices
given, we have $d = \order(1)$ and $\degpoly = \order(\log
\matdim)$. In order to make
inequality~\eqref{eqn:rand-alg-bound} hold with probability at least
$1-\epsilon$, the upper bound~\eqref{eqn:S-lower-upper} suggests
choosing $\NUMREP = \Theta(\log(1/\epsilon))$.

Finally, we need to upper bound the quantity $\order(\log(1/\tau))$.
In order to do so, let us revisit Algorithm~\ref{alg:rand-rank-esti}
to see how rounding errors affect the final output.  For each integer
$k=1, \ldots, 2 \degpoly+1$, let us denote by $\delta_k$ the error of
evaluating $q_1^k(A) \gaussvec$ using Algorithm~\ref{alg:poly-eval}.
It is known~\cite[][Chapter 2.4.2]{mason2010chebyshev} that the
rounding error of evaluating a Chebyshev expansion is bounded by
$\nummac d \tau$. Thus, we have $\delta_{k+1} \leq \ltwos{q_1(A)}
\delta_k + \nummac d \tau$.  Since $\ltwos{q_1(A)} \leq 1.1$ by
construction, we have the upper bound
\begin{align}
\label{EqnIntermediate}
\delta_k & \leq 10 (1.1^{k+1}-1) \nummac d \tau.
\end{align}

For a polynomial of the form $q_2(x) = \sum_{i=0}^{2 \degpoly + 1} a_i
x^i$, we have $y = \sum_{i=0}^{2 \degpoly + 1} a_i q_1^i(A) b$.  As a
consequence, there is a universal constant $C$ such that error in
evaluating $y$ is bounded by
\begin{align*}
C \sum_{i=0}^{2 \degpoly + 1} \delta_i |a_i| & \leq C' \; (1.1)^{2
  \degpoly + 1} \nummac d \tau \sum_{i=0}^{2 \degpoly +1} |a_i|.
\end{align*}
By the definition of the polynomial $q_2$ and the binomial theorem, we
have
\begin{align*}
\sum_{i=0}^{2 \degpoly + 1} |a_i| \leq \frac{2^ \degpoly}{B(\degpoly +
  1, \degpoly + 1)} = \frac{2^\degpoly (2 \degpoly +
  1)!}{(\degpoly!)^2} \leq 2^{3 \degpoly}.
\end{align*}
Putting the pieces together, in order to make the overall error small, it
suffices to choose $\tau$ of the order $(\nummac d \matdim)^{-1} 2^{-4
  \degpoly}$.  Doing so ensures that $\log(1/\tau) = \order(\degpoly
\log(\nummac d \matdim))$, which when combined with our earlier upper
bounds on $d$, $\degpoly$ and $\NUMREP$, establishes the
claim~\eqref{EqnRandUpper}.


\subsection{Proof of Theorem~\ref{ThmRandLower}}
\label{SecProofThmRandLower}

In order to prove Theorem~\ref{ThmRandLower}, it suffices to consider
the two-player setting, since the first two machines can always
simulate the two players Alice and Bob.  Our proof proceeds via
reduction from the 2-SUM problem~\cite{woodruff2014optimal}, in which
Alice and Bob have inputs $(U_1,\dots,U_\matrank)$ and
$(V_1,\dots,V_\matrank)$, where each $U_i$ and $V_i$ are subsets of
$\{1,\ldots, L\}$. It is promised that for every index $i \in \{1,
\ldots, \matrank\}$, the intersection of $U_i$ and $V_i$ contains at
most one element.  The goal is to compute the sum $\sum_{i=1}^\matrank
|U_i\cap V_i|$ up to an additive error of
$\sqrt{\matrank}/2$. Woodruff and Zhang~\cite{woodruff2014optimal}
showed that randomized communication complexity of the 2-SUM problem
is lower bounded as $\Omega(\matrank L)$.

We note here that when $\matrank \geq 16$, the same communication
complexity lower bound holds if we allow the additive error to be $2 
\sqrt{\matrank}$. To see this, suppose that Alice and Bob have inputs
of length $\matrank/16$ instead of $\matrank$.  By replicating their
inputs $16$ times, each of Alice and Bob can begin with an input of
length $\matrank$.  Assume that by using some algorithm, they can
compute the 2-SUM for the replicated input with additive error at most
$2 \sqrt{\matrank}$.  In this way, they have computed the 2-SUM for
the original input with additive error at most
$\sqrt{\matrank}/8$. Note that $\sqrt{\matrank/8} =
\sqrt{\matrank/16}/2$. The lower bound on the 2-SUM problem implies
that the communication cost of the algorithm is $\Omega(\matrank
L/16)$, which is on the same order of $\Omega(\matrank L)$.

To perform the reduction, let $L = \lfloor \matdim/\matrank- 1
\rfloor$. Since $\matrank \leq \matdim/2$, we have $L \geq 1$.
Suppose that Alice and Bob are given subsets $(U_1,\dots,U_\matrank)$
and $(V_1,\dots,V_\matrank)$, which define an underlying instance of
the 2-SUM problem.  Based on these subsets, we construct two
$\matdim$-dimensional matrices $A_1$ and $A_2$ and the matrix sum $A
\defeq A_1 + A_2$; we then argue that any algorithm that can estimate
the generalized matrix rank of $A$ can solve the underlying 2-SUM
problem. 

The reduction consists of the following steps.  First, Alice
constructs a matrix $X$ of dimensions $\matrank L \times n$ as follows.
For each $i \in \{1,\dots,\matrank\}$ and $j \in \{1,\dots,L\}$,
define $t(i,j) = (i-1)L+j$, and let $X_{t(i,j)}$ denote the associated
row of $X$.  Letting $e_{t(i,j)} \in \real^\matdim$ denote the
canonical basis vector (with a single one in entry $t(i,j)$), we
define
\begin{align*}
X_{t(i,j)} & = \begin{cases} e_{t(i,j)} & \mbox{if $j \in U_i$} \\ 0 &
  \mbox{otherwise.}
\end{cases}
\end{align*}
Second, Bob constructs a matrix $Y$ of dimensions $\matrank L \times n$
following the same rule as Alice, but using the subset $(V_1, \ldots,
V_L)$ in place of $(U_1, \ldots, U_L)$.  Now define the $\matdim
\times \matdim$ matrices
\begin{align*}
A_1 \defeq \ccontwo \Big(X^T X + \sum_{i=1}^\matrank e_{\matrank L+i}
e_{\matrank L+i}^T\Big) \quad \mbox{and} \quad A_2 \defeq \ccontwo
\Big(Y^T Y + \sum_{i=1}^\matrank e_{\matrank L+i} e_{ \matrank
  L+i}^T\Big).
\end{align*}
With these definitions, it can be verified that $\ltwos{A}\leq
2\ccontwo\leq 1$, and moreover that all eigenvalues of $A$ are either
equal to $2\ccontwo$ or at most $\ccontwo$. Since $\cconone <
2\ccontwo$, the quantities $\rank(A,\cconone)$ and $\rank(A,\ccontwo)$
are equal, and equal to the number of eigenvalues at $2 \ccontwo$.
The second term in the definition of $A_1$ and $A_2$ ensures that
there are at least $\matrank$ eigenvalues equal to $2\ccontwo$.  For
all $(i,j)$ pairs such that $j\in U_i\cap V_i$, the construction of
$X$ and $Y$ implies that there are two corresponding rows in $X$ and
$Y$ equal to each other, and both of them are canonical basis vectors.
Consequently, they create a $2 \ccontwo$ eigenvalue in matrix
$A$. Overall, we have $\rank(A,\cconone) = \rank(A,\ccontwo) =
\matrank + \sum_{i=1}^\matrank |U_i\cap V_i|$, Since the problem
set-up ensures that $|U_i\cap V_i| \leq 1$, we conclude $ \matrank
\leq \rank(A,\cconone) \leq 2 \matrank$. \\

Now suppose that there is a randomized algorithm estimating the rank
of $A$ such that
\begin{align*}
(1-\delta) \rank(A,\cconone) \leq \rhat(A) \leq (1+\delta)
  \rank(A,\ccontwo).
\end{align*}
Introducing the shorthand $s \defeq \sum_{i=1}^k |U_i\cap V_i|$, when $\delta =
1/\sqrt{\matrank}$, we have
  \begin{align*}
\matrank + s - (\matrank + s)/\sqrt{\matrank} \leq \rhat(A) \leq
\matrank + s + (\matrank + s)/\sqrt{\matrank}.
\end{align*}
Thus, the estimator $\rhat(A) - \matrank $ computes $s$ up to additive
error $(\matrank + s)/\sqrt{\matrank}$, which is upper bounded by
$2\sqrt{\matrank}$. It means that the rank estimation algorithm solves
the 2-SUM problem.  As a consequence, the randomized communication
complexity of the rank estimation problem is lower bounded
by~$\Omega(\matrank L) = \Omega(n)$.


\section{Discussion}

In this paper, we have studied the problem of estimating the
generalized rank of matrices.  Our main results are to show that in
the deterministic setting, sending $\Theta(n^2)$ bits is both
necessary and sufficient in order to obtain any constant relative
error.  In contrast, when randomized algorithms are allowed, this
scaling is reduced to $\widetilde{\Theta}(n)$.

Our work suggests an important problem, one whose resolution has a
number of interesting consequences.  In the current paper, we
establish the $\widetilde\Theta(n)$ scaling of communication
complexity for achieving a relative error $\delta = 1/\sqrt{\matrank}$
where $\matrank$ is the matrix rank.  Moreover,
Algorithm~\ref{alg:rand-rank-esti} does not guarantee higher
accuracies (e.g., $\delta = 1/\matrank$), and as discussed in
Section~\ref{SecRandLower}, it is unknown whether the $\Omega(n)$
lower bound is tight.  The same question remains open even for the
special case when all the matrix eigenvalues are either greater than
constant $c$ or equal to zero.  In this special case, if we were to
set $c_1 = c$ and $c_2 = 0$ in Algorithm~\ref{alg:rand-rank-esti},
then it would compute \emph{ordinary} matrix rank with relative error
$\delta = 1/\sqrt{\matrank}$.  Although the problem is easier in the
sense that all eigenvalue are promised to lie in the subset $\{0\}\cup
(c,1]$, we are currently not aware of any algorithm with
$\widetilde\order(n)$ communication cost achieving better error rate.
On the other hand, proving a tight lower bound for arbitrary $\delta$
remains an open problem.

The special case described above is of fundamental interest because it
can be reduced to many classical problems in linear algebra and convex
optimization, as we describe here.  More precisely, if there is an
algorithm solving any of these problems, then it can be used for
computing the matrix rank with relative error $\delta = 0$.  On the
other hand, if we obtain a tight lower bound for computing the matrix
rank, then it implies a lower bound for a larger family of problems. We
list a subset of these problems giving a rough intuition for the reduction.

To understand the connection, we begin by observing that the problem of
rank computation can be reduced to that of matrix rank testing, in
which the goal is to determine whether a given matrix sum $A \defeq
A_1 + \dots + A_m$ has rank at most $\matrank-1$, or rank at least
$\matrank$, assuming that all eigenvalues belong to $\{0\} \cup
(c,+\infty)$. If there is an algorithm solving this problem for
arbitrary integer $\matrank \leq \matdim$, then we can use it for
computing the rank.  The reduction is by performing a series of binary
searches, each step deciding whether the rank is above or below a
threshold. In turn, the rank test problem can be further reduced to
the following problems:

\vspace{-10pt}
\paragraph{Singularity testing:} 
The goal of singularity testing is to determine if the sum of matrices
$B\defeq B_1+\dots+B_m$ is singular, where machine $i$ stores the PSD
matrix $B_i$. Algorithms for singularity testing can be used for rank
testing. The reduction is by using a public random coin to generate a
shared random projection matrix $Q\in \R^{\matrank \times n}$ on each
machine and then setting $B_i \defeq Q A_i Q^T$. The inclusion of the
public coin only increases the communication complexity by a moderate
amount~\cite{lee2009lower}, in particular by an additive term
$\order(\log(n))$.  On the other hand, with high probability the
matrix $A$ has rank at most $\matrank-1$ if and only if the matrix $B$
is singular.

\vspace{-10pt}
\paragraph{Solving linear equations:} Now suppose that machine $i$
stores a strictly positive definite matrix $C_i$ and a vector $y$. The
goal is to compute the vector $x$ satisfying $C x = y$ for $C\defeq
C_1 + \dots+C_m$. Algorithms for solving linear equations can be used
for the singularity test. In particular, let $C_i \defeq B_i + \lambda I$
and take $y$ to be a random Gaussian vector. If the matrix $B$ is
singular, then the norm $\ltwos{x}\to \infty$ as $\lambda\to
0$. Otherwise, it remains finite as $\lambda\to 0$. Thus, we can test
for $\lambda = 1,\frac{1}{2},\frac{1}{4},\frac{1}{8},\dots$ to decide
if the matrix is singular. Note that the solution need not be exact,
since we only test if the $\ell_2$-norm remains finite.

\vspace{-10pt}
\paragraph{Convex optimization:} Suppose that each machine has a 
strictly convex function $f_i$, and the overall goal is to compute a
vector $x$ that minimizes the function $x \mapsto f(x) \defeq f_1(x) +
\cdots + f_m(x)$. The algorithms solving this problem can be used for
solving linear equations. In particular, for a strictly positive
definite matrix $C_i$, the function $f_i(x) \defeq \frac{1}{2} x^T C_i
x - \frac{1}{m} y^T x$ is strictly convex, and with these chocies, the
function $f$ is uniquely minimized at $C^{-1} y$. (Since the linear
equation solver doesn't need to be exact, the solution here is also
allowed to be approximate.)\\

\vspace{-5pt} This reduction chain suggests the importance of studying
matrix rank estimation, especially for characterizing lower bounds 
on communication complexity.  We hope the results in this paper are 
a meaningful first step in exploring this problem area.


\subsection*{Acknowledgements}

MJW and YZ were partially supported by the ONR-MURI grant DOD 002888 from the Office of Naval Research. MIJ and YZ were partially supported by the U. S. Army Research Laboratory and the U. S. Army Research Office under contract/grant number W911NF-11-1-0391. In addition, YZ was partially supported by a Baidu
Fellowship.

\appendix


\section{Proof of Lemma~\ref{lemma:poly-approx}}
\label{sec:proof-poly-approx}

The function $q_2$ is monotonically increasing on $[0,1]$.  In
addition, we have $q_2(0)=0$ and $q_2(1) = 1$, and hence $q_2(z) \in
[0,1]$ for all $z \in [0,1]$.  Let us refine this analysis on two end
intervals: namely, $z \in [-0.1,0.1]$ and $z\in [0.9,1.1]$.  For $z
\in [-0.1, 0.1]$, it is easy to observe from the definition of $q_2$
that $q_2(z) \geq 0$. Moreover, for $z\in [-0.1,0.1]$ we have
$|z(1-z)| \leq 0.11$. Thus,
\begin{align*}
q_2(z) & = \frac{\int_0^z t^{p } (1-t)^{p } d t}{\int_0^1 t^{p }
  (1-t)^{p } d t} \leq \frac{\int_0^z t^{p } (1-t)^{p } d
  t}{\int_{0.4}^{0.6} t^{p } (1-t)^{p } d t} \leq \frac{0.1 \times
  (0.11)^{p }}{0.2 \times (0.24)^{p }} < 2^{-p }.
\end{align*}
The function $q_2$ is symmetric in the sense that $q_2(z) + q_2(1-z) =
1$. Thus, for $z\in [0.9,1.1]$, we have $q_2(z) = 1 - q_2(1 - z) \in
[1 - 2^{-p},1]$.  In summary, we have proved that
\begin{subequations}
\begin{align}
&0\leq q_2(z) \leq 1 \mbox{ for } z\in [-0.1,1.1],\label{eqn:q2-bound-1}\\
&q_2(z) \leq 2^{-p} \mbox{ for } z\in [-0.1,0.1],\label{eqn:q2-bound-2}\\
&q_2(z) \geq 1-2^{-p} \mbox{ for } z\in[0.9,1.1].\label{eqn:q2-bound-3}
\end{align}
\end{subequations}

By the standard uniform Chebyshev approximation, we are guaranteed
that $q_1(x)\in [-0.1,1.1]$ for all $x \in [0,1]$.  Thus,
inequality~\eqref{eqn:q2-bound-1} implies that $q_2(q_1(x)) \in [0,1]$
for all $x \in [0,1]$.  If $x \in [0,\ccontwo]$, then $q_1(x)\in
[-0.1,0.1]$, and thus inequality~\eqref{eqn:q2-bound-2} implies
$q_2(q_1(x))\leq 2^{-p}$.  If $x\in [\cconone,1]$, then $q_1(x)\in
[0.9,1.1]$, and thus inequality~\eqref{eqn:q2-bound-3} implies
$q_2(q_1(x))\geq 1 - 2^{-p}$.  Combining the last two inequalities
yields that
\begin{align*}
|q_2(q_1(x)) - \HSPEC(x)|\leq 2^{-p} \qquad \mbox{for all $x\in
  [0,\ccontwo] \cup [\cconone,1]$.}
\end{align*}


\section{Proof of Lemma~\ref{lemma:intersec-ortho-space}}
\label{sec:proof-intersec-ortho-space}

Let $q_t$ be the $t$-th row of $Q_2$, and let $Q^{(t)} \in \R^{\matrank +t}$ be
the matrix whose first $\matrank $ rows are the rows of $Q_1$, and its
remaining $t$ rows are $q_1,\ldots,q_t$. Let $q_{t+1}^{\parallel}$ be
the projection of $q_{t+1}$ to the subspace generated by the rows of
$Q^{(t)}$ and let $q_{t+1}^{\perp} \defeq q_{t+1} -
q_{t+1}^{\parallel}$.  We have
\begin{align*}
(Q^{(t+1)})^T Q^{(t+1)} &= (Q^{(t)})^T Q^{(t)} + q_t^T q_t = (Q^{(t)})^T Q^{(t)} +
(q_{t+1}^{\parallel})^T q_{t+1}^{\parallel} +
(q_{t+1}^{\perp})^Tq_{t+1}^{\perp} \\
&\succeq (Q^{(t)})^T Q^{(t)} +
(q_{t+1}^{\perp})^Tq_{t+1}^{\perp}.
\end{align*}
This inequality yields the lower bound
\begin{align}
\label{eqn:rand-matrix-decompose}
Q_1^T Q_1 + Q_2^T Q_2 \succeq Q_1^T Q_1 + \sum_{t=1}^{\matrank }
(q_{t}^{\perp})^T q_{t}^{\perp},
\end{align}
where $\succeq$ denotes ordering in the positive semidefinite cone.
Note that the rows of $Q_1$ and $\{q_t^{\perp}\}_{t=1}^{\matrank }$ are
mutually orthogonal. To prove that the $\frac{6k}{5}$-th largest
eigenvalue of $Q_1^T Q_1 + Q_2^T Q_2$ is greater than $1/10$, it
suffices to prove that there are at least $\matrank /5$ vectors in
$\{q_t^{\perp}\}_{t=1}^{\matrank }$ which satisfy $\ltwos{q_{t}^{\perp}}^2 >
1/10$.

Let $\mathcal{S}_1$ be the linear subspace generated by
$q_1,\dots,q_{t-1}$ and let $\mathcal{S}_1^\perp$ be its orthogonal
subspace. The vector $q_t$ is uniformly sampled from a unit sphere in
$\mathcal{S}_1^\perp$. Let  $\mathcal{S}_2$  be the linear subspace
generated by the rows of $Q^{(t-1)}$. Since $Q^{(t-1)}$ has $\matrank +t-1$ rows, the
subspace has at most $\matrank +t-1$ dimensions. Without loss of generality,
we assume that $\mathcal{S}_2$ has $\matrank +t-1$ dimensions (otherwise, we
expand it to reach the desired dimensionality). We let
$\mathcal{S}_2^\perp$ be the orthogonal subspace of
$\mathcal{S}_2$. By definition, $q_{t}^{\perp}$ is the projection of
$q_t$ to $\mathcal{S}_2^\perp$ (or a linear space that contains
$\mathcal{S}_2^\perp$ if the subspace $\mathcal{S}_2$ has been expanded to reach the
$\matrank +t-1$ dimensionality). Let $q'_{t}$ be the projection of $q_t$ to
$\mathcal{S}_1^\perp\cap\mathcal{S}_2^\perp$, then we have
\begin{align}
\label{eqn:proj-and-subproj-relation} 
\ltwos{q_{t}^{\perp}}^2 \geq \ltwos{q'_{t}}^2.
\end{align}
Note that $\mathcal{S}_1^\perp$ is of dimension $n-t+1$ and
$\mathcal{S}_2^\perp$ is of dimension $n-\matrank -t+1$. Thus, the dimension
of $\mathcal{S}_1^\perp\cap\mathcal{S}_2^\perp$ is at least
$n-\matrank -2t+2$. Constructing $q'_{t}$ is equivalent to projecting a random
vector in the $(n-t+1)$-dimension sphere to a $(n-\matrank -2t+2)$-dimension
subspace.  It is a standard result~(e.g.~\cite[][Lemma
  2.2]{dasgupta2003elementary}) that
\begin{align*}
\prob\Big[ \ltwos{q_t'}^2 \leq \beta\cdot \frac{n-\matrank -2t+2}{n-t+1} \Big]
\leq \exp\Big(\frac{n-\matrank -2t+2}{2}(1-\beta+\log(\beta))\Big) \quad
\mbox{for any $\beta < 1$}.
\end{align*}
Setting $\beta = 0.3$ and using the fact that $t\leq \matrank  \leq n/4$, we
find that
\begin{align}
\label{eqn:aug-dim-bound}
\prob\Big[ \ltwos{q_t'}^2 \leq 1/10 \Big] \leq
\exp\Big(\frac{n-n/4-n/2+2}{2}(1-0.3+\log(0.3))\Big) \leq \exp(-n/16).
\end{align}
Defining the event $\event_t \defeq \{ \ltwos{q'_{t}}^2 \leq 1/10 \}$,
note that inequality~\eqref{eqn:aug-dim-bound} yields $\prob[\event_t]
\leq \exp(-n/16)$. Since $q_t'$ is the projection of a random unit
vector to a subspace of constant dimension, the events
$\{\event_j\}_{j=1}^t$ are mutually independent, and hence
\begin{align*}
\prob\Big[ \mbox{at least $\frac{4k}{5}$ events in
    $\{\Event_j\}_{j=1}^t$ occur } \Big] \leq {\matrank  \choose 4\matrank/5}
(\exp(-n/16))^{\frac{4\matrank}{5}} & \leq \exp\Big(\frac{\matrank \log(\matrank )}{5} -
\frac{\matrank  n}{20}\Big) \\
& \leq \exp\Big(- \frac{3 \matrank  n}{100}\Big),
\end{align*}
where the last inequality follows since any integer $\matrank $ satisfies
$\log(\matrank ) \leq \frac{2\matrank}{5} \leq \frac{n}{10}$.  Thus, with probability
at least $1 - \exp(-\frac{3 \matrank  n}{100})$, there are at least $\matrank/5$ rows
satisfying $\ltwos{q'_{t}}^2 > 1/10$. Combining this result with
inequality~\eqref{eqn:rand-matrix-decompose}
and~\eqref{eqn:proj-and-subproj-relation} completes the proof.


\section{Proof of Lemma~\ref{lemma:sub-exponential}}
\label{sec:proof-sub-exponential}

The claimed facts about sub-exponential random variables are
standard~\cite{BulKoz}, but we provide proofs here for completeness.

\paragraph{Part (a):} 
Let $Z$ be $\chi^2$ variable with one degree of freedom.  Its moment
generating function takes the form
\begin{align*}
\E[\exp(t(Z-1))] = (1-2t)^{-1/2} e^{-t} \quad \mbox{for $t< 1/2$}.
\end{align*}
Some elementary algebra shows that $(1-2t)^{-1/2} e^{-t} \leq
e^{2t^2}$ for any $t \in [-1/4,1/4]$. Thus, we have $\E[\exp(t(Z-1))]
\leq e^{2t^2}$ for $|t|\leq 1/4$, verifying the recentered variable $X
= Z-1$ is sub-exponential with parameter $(4,4)$. Also by the moment
generating function of $Z$, we have
\begin{align*}	
\E[\exp(t(1-Z))] = (1+2t)^{-1/2} e^{t} \quad \mbox{for $t > -1/2$}.
\end{align*}
Replacing $t$ by $-t$ and comparing with the previous conclusion
reveals that $1-Z$ is sub-exponential with parameter $(4,4)$.

\paragraph{Part (b):} Suppose that $Z_1,\dots,Z_n$ are independent and $Z_i$ is 
sub-exponential with parameter $(\sigma_i^2,\beta_i)$.  By the
definition of sub-exponential random variable, we have
\begin{align*}
\E\Big[\exp\Big(t\sum_{i=1}^n \alpha_i Z_i\Big)\Big] = \prod_{i=1}^n
\E[\exp(t \alpha_i Z_i)] \leq \prod_{i=1}^n \exp((t
\alpha_i)^2\sigma^2/2) = \exp\Big( \frac{t^2 \sum_{i=1}^n \alpha_i^2
  \sigma_i^2}{2} \Big)
\end{align*}
for all $t\leq \max_{i\in[n]}\{1/(\alpha_i\beta_i)\}$. This bound
establishes that $\sum_{i=1}^n \alpha_i Z_i$ is sub-exponential with
parameter $(\sum_{i=1}^n
\alpha_i^2\sigma_i^2,\max_{i\in[n]}\{\alpha_i\beta_i\})$, as claimed.

\paragraph{Part (c):} 
Notice that $\mprob[Z \geq t] = \mprob[ e^{\lambda Z} \geq e^{\lambda
    t}]$ with any $\lambda > 0$. Applying Markov's inequality yields
\begin{align*}
\mprob[Z \geq t] \leq \frac{\E[\exp(\lambda Z)]}{e^{\lambda t}} \leq
\exp\Big( -\lambda t + \frac{\lambda^2\sigma^2}{2}\Big) \quad
\mbox{for $\lambda \leq 1/\beta$},
\end{align*}
where the last step follows since $Z$ is sub-exponential with
parameter $(\sigma^2,\beta)$.  Notice that the minimum of $-\lambda t
+ \frac{\lambda^2\sigma^2}{2}$ occurs when $\lambda^* = t/\sigma^2$.
Since $t < \sigma^2/\beta$, we have $\lambda^* < 1/\beta$, verifying
the validness of $\lambda^*$.  Plugging $\lambda^*$ in the previous
inequality completes the proof.


\bibliographystyle{abbrvnat} \bibliography{bib}


\end{document}